\documentclass[%
 reprint,
superscriptaddress,
 amsmath,amssymb,
 aps,
 prfluids,
]{revtex4-2}

\usepackage{graphicx}
\usepackage{dcolumn}
\usepackage{bm}
\usepackage{hyperref}
\usepackage{xcolor}
\usepackage{siunitx}
\usepackage{csquotes}
\usepackage{placeins} 

\newcommand{\iu}{\mathrm{i}\mkern1mu}

\begin{document}

\title{Sound waves, diffusive transport, and wall slip in nanoconfined compressible fluids}

\author{Hannes Holey}
\email{hannes.holey@kit.edu}
\affiliation{%
Institute for Applied Materials, Karlsruhe Institute of Technology,
Straße am Forum 7, 76131 Karlsruhe, Germany
}%
\affiliation{%
Department of Microsystems Engineering (IMTEK), University of Freiburg, 
Georges-Köhler-Allee 103, 79110 Freiburg, Germany
}%

\author{Peter Gumbsch}
\affiliation{%
Institute for Applied Materials, Karlsruhe Institute of Technology, 
Straße am Forum 7, 76131 Karlsruhe, Germany
}%
\affiliation{%
Fraunhofer Institute for Mechanics of Materials IWM, 
Wöhlerstraße 11, 79108 Freiburg, Germany
}%

\author{Lars Pastewka}
\email{lars.pastewka@imtek.uni-freiburg.de}
\affiliation{%
Department of Microsystems Engineering (IMTEK), University of Freiburg, 
Georges-Köhler-Allee 103, 79110 Freiburg, Germany
}%
\affiliation{%
Cluster of Excellence livMatS, 
Freiburg Center for Interactive Materials and Bioinspired Technologies, 
University of Freiburg, Georges-Koehler-Allee 105, 79110 Freiburg, Germany
}

\date{\today}

\begin{abstract}
Although continuum theories have been proven quite robust to describe confined
fluid flow at molecular length scales, molecular dynamics (MD) simulations
reveal mechanistic insights into the interfacial dissipation processes.
Most MD simulations of confined fluids have used setups in which the lateral box
size is not much larger than the gap height, thus breaking thin-film
assumptions usually employed in continuum simulations.
Here, we explicitly probe the long wavelength hydrodynamic correlations in
confined simple fluids with MD and compare to gap-averaged continuum theories
as typically applied in e.\,g. lubrication.
Relaxation times obtained from equilibrium fluctuations interpolate between the
theoretical limits from bulk hydrodynamics and continuum formulations with
increasing wavelength.
We show how to exploit this characteristic transition to measure viscosity and
slip length in confined systems simultaneously from equilibrium MD
simulations.
Moreover, the gap-averaged theory describes a geometry-induced dispersion
relation that leads to overdamped sound relaxation at large wavelengths, which
is confirmed by our MD simulations.
Our results add to the understanding of transport processes under strong
confinement and might be of technological relevance for the design of
nanofluidic devices due to the recent progress in fabrication methods.
\end{abstract}

\maketitle

\section{Introduction}
\label{sec:intro}

The field of nanofluidics has emerged from microfluidics within the last 20
years due to progress in nanofabrication, characterization and simulation
tools \citep{schoch2008_transport, sparreboom2009_principles,
bocquet2010_nanofluidics}.
The reduction of characteristic length scales led to new phenomena that occur
with increasing surface-to-volume ratios.
A primary application of nanofluidic systems is single molecule selectivity in
biotechnological lab-on-a-chip devices \citep
{mijatovic2005_technologies,abgrall2008_nanofluidic}.
The ultimate lower length scale limit of continuum theories has been determined
to be $1\,\mathrm{nm}$, and indeed, many effects occur on length scales well
above this limit \citep{bocquet2010_nanofluidics}.
However, recent advances in the fabrication of devices with sub-nanometer
confinement \citep{feng2016_singlelayer, radha2016_molecular,
tunuguntla2017_enhanced} require theories that consider the discrete nature of
molecules, and particularly, interactions between fluid molecules and the
electronic structure of the confining walls become relevant \citep
{kavokine2021_fluids, kavokine2022_fluctuationinduced}.

The transition from continuum to molecular descriptions of confined fluids is
closely related to fluid structuring.
Molecular arrangement into distinct layers occurs due to geometric confinement
and is strongly influenced by the interaction between the fluid and wall
atoms \citep{israelachvili1983_molecular, chan1985_drainage,
israelachvili1988_forces}.
Molecular dynamics (MD) simulations have been widely employed to study the
effect of fluid structure on transport properties \citep
{thompson1992_phase, gao1997_layering, gao1997_origins, gao1997_structure,
jabbarzadeh1997_rheological} and the fluid--wall boundary condition \citep
{thompson1990_shear, thompson1997_general, cieplak2001_boundary}.
While MD simulations are a valuable tool to study confined system, \citet
{travis1997_departure} showed that classical continuum theories remain valid
down to about five molecular diameters in simple fluids.
Therefore, dimensionality-reduced continuum formulations are often employed on
these scales, and combined with MD parameterizations, e.\,g. for wall
slip \citep{savio2015_multiscale}.

In many continuum simulations of dense confined fluids, compressibility effects
are neglected, which appears reasonable at first glance.
Sound speeds in liquids are on the order of $10^3\,\mathrm{m/s}$, and thus plays
a minor role in momentum transport at hydrodynamic length and time
scales \citep{alder1970_decay}.
However, in the presence of walls, friction strongly influences the dynamics of
sound waves, as was first pointed out by \citet{ramaswamy1982_linear} in a
phenomenological approach for adsorbed layers on substrates.
In their work, an unspecified friction term was used to interpolate between
compressible hydrodynamics and Fickian diffusion.
The effect of overdamped sound was later related to the (negative) algebraic
long-time tail of the velocity autocorrelation function of a suspended particle
in a confined system observed in lattice Boltzmann simulations \citep
{hagen1997_algebraic, pagonabarraga1998_algebraic}.
Although the diffusive sound modes govern the velocity of the suspended particle
at long times, they do not contribute to the diffusion coefficient.
A rigorous mathematical treatment of the problem was subsequently given in a
series of works by Felderhof for a single plane wall~\citep
{felderhof2005_effect}, two parallel plane walls ~\citep
{felderhof2006_diffusion}, and circular geometries \citep
{felderhof2010_transient}.

It is common practice to extract transport coefficients from the correlations of
equilibrium fluctuations, e.\,g. through the Green-Kubo approach \citep
{green1954_markoff, kubo1957_statisticalmechanical}, or by means of a direct
fit to the autocorrelation functions of hydrodynamic variables \citep
{palmer1994_transversecurrent, cheng2020_computing}.

Here, we briefly recap bulk hydrodynamic theory in order to motivate a similar
approach for confined fluids, see Refs.~\citep
{boon1980_molecular, hansen2007_theory} for more details.
We start from the Navier-Stokes equations that describe mass, momentum, and
energy balance in terms of the density $\rho(\vec{r})$, momentum $\vec{j}(\vec
{r})$ and energy $e(\vec{r})$ fields.
Here and in the following, $\vec{r}=(x,y,z)$ denotes a spatial coordinate.
Given the trajectory $\{\vec{r}_i(t),\,\vec{v}_i(t)\}$ of $N$ particles, we
compute the mass density field as
\begin{equation}
    \rho(\vec{r}, t) = m \sum_{i = 1}^N \delta[\vec{r} - \vec{r}_i(t)],
\end{equation} and the corresponding Fourier transform is given by
\begin{equation}
\begin{split}
    \tilde{\rho}(\vec{k}, t) &= \frac{1}{V}\int_V \mathrm{d}\vec{r}\rho(\vec
     {r}, t)\exp(-i\vec{k}\cdot\vec{r})\\
    &= \frac{m}{V}\sum_{i = 1}^N\exp(-i\vec{k}\cdot\vec{r}_i(t)).
\end{split}
\end{equation}
Similarly, we obtain for the Fourier coefficients of the momentum density
\begin{equation}
    \tilde{\vec{j}}(\vec{k}, t)= \frac{m}{V}\sum_{i = 1}^N\vec{v}_i(t)\exp
     (-i\vec{k}\cdot\vec{r}_i(t)).
\end{equation}
For small deviations from equilibrium $\delta\rho(\vec{k},t) = \rho(\vec
{k},t)-\langle\rho(\vec{k},t)\rangle$, the linearized hydrodynamic equations
can be solved.
For the sake of brevity, we omit the $\delta$-notation from now on.
The time evolution is typically solved through a Laplace transform, and one
finds that the longitudinal and transverse momentum modes decouple.
The normalized time autocorrelation function of the Fourier coefficients of
transverse momentum decays exponentially
\begin{equation}\label{eq:acf_transverse} C_\perp(k,t) \equiv \frac
 {\langle j_\perp^\ast(k,0) j_\perp(k,t) \rangle}{\langle j_\perp^\ast
 (k,0) j_\perp(k,0) \rangle} = \exp(-\nu k^2 t),
\end{equation}
where $\vec{j}_\perp = \vec{j} - j_\parallel \hat{\vec{k}}$ and $j_\parallel
= \vec{j}\cdot \hat{\vec{k}}$ with $\hat{\vec{k}} = \vec{k}/|\vec{k}|$ are the
momentum fluxes perpendicular and parallel to the wavevector $\vec
{k}$, respectively.
The angular brackets denote an average over initial conditions, and the star is
the complex conjugate.
The decay rate is proportional to the kinematic viscosity $\nu=\eta/\rho$, where
$\eta$ is the conventional shear viscosity.
The kinematic viscosity has units of a diffusion constant, and can be regarded
as such for the diffusion of momentum.

For the inverse Laplace transform of the longitudinal modes, one usually uses a
second order approximation in $k=|\vec{k}|$ (instead of finding the exact roots
of a cubic equation) \citep{mountain1966_spectral}.
The normalized longitudinal momentum autocorrelation function has propagating
sound modes and the decay rate is determined by viscous and conductive effects
\begin{equation}\label{eq:acf_longitudinal} C_\parallel(k,t) \equiv \frac
 {\langle j_\parallel^\ast(k,0) j_\parallel(k,t)\rangle}
 {\langle j_\parallel^\ast(k,0) j_\parallel(k,0)\rangle} = \exp(-\Gamma k^2
 t)\cos(c_\mathrm{s} k t),
\end{equation}
with sound attenuation coefficient $\Gamma=(\gamma-1)D_\mathrm
{T}/2 + \nu_\mathrm{L}/2$ and adiabatic speed of sound $c_\mathrm{s}$. Here,
$\gamma=c_\mathrm{P}/c_\mathrm{V}$ denotes the ratio of volume-specific
isobaric and isochoric heat capacities, $D_\mathrm{T}=\kappa_\mathrm
{T}/c_\mathrm{P}$ is the thermal diffusivity with thermal conductivity
$\kappa_\mathrm{T}$, and $\nu_\mathrm{L}=(4\eta/3 + \zeta)/\rho$ is the
kinematic longitudinal viscosity with bulk viscosity $\zeta$.
Finally, the normalized density autocorrelation function is given by
\begin{equation}\label{eq:acf_density}
    \begin{split} C_\rho(k,t) \equiv& \frac{\langle\rho^\ast(k,0) \rho
     (k,t) \rangle}{\langle \rho^\ast(k,0) \rho(k,0)\rangle} \\ =& \frac
     {\gamma-1}{\gamma}e^{-D_\mathrm{T}k^2t}
     + \frac{1}{\gamma}e^{-\Gamma k^2 t}
    \cos(c_\mathrm{s} k t),
    \end{split}
\end{equation}
where the first and second term describe the Rayleigh and Brillouin process,
respectively.
Equation.~\eqref{eq:acf_density} has probably received most attention, since its
power spectrum---the dynamic structure factor $S
(k, \omega)$---is experimentally accessible, for instance through light
scattering \citep{berne2000_dynamic}.
The set of correlation functions fully describes the dynamics of statistically
independent fluctuating quantities.
The fluctuations of density are assumed to occur on time scales that do not
allow the exchange of heat.
Hence, the adiabatic compressibility governs the velocity of sound ($D_\mathrm
{T}k\ll c_\mathrm{s}$) \citep{boon1980_molecular}.

In this paper, we investigate the time correlations of confined systems.
\citet{gutkowicz-krusin1982_equilibrium, gutkowicz-krusin1983_effects} have
 derived the dynamic structure factor for confined fluids, considering both
 momentum and energy transport at the fluid-wall interface.
However, their approach considers only no-slip or perfect slip boundary
conditions.
\citet{bocquet1993_hydrodynamic, bocquet1994_hydrodynamic} derived momentum time
 correlation functions for confined systems, and related these to the
 hydrodynamic boundary conditions with partial-slip.
Yet, they considered only transverse fluxes perpendicular to the walls, averaged
over the lateral dimensions of the system.
The effect of the lateral periodic box size in MD simulations of confined
systems, i.\,e. the admissible wavelengths of longitudinal and in-plane
transverse modes, have received little attention so far \citep
{ogawa2019_large}.

Here, we provide a derivation of time correlation functions for height-averaged
balance equations---similar to the continuum methods commonly applied in
lubrication \citep{szeri1998_fluid}---under isothermal conditions.
Thus, we consider only momentum transport at the interface with partial-slip
boundary conditions, which allows us to extract effective transport
coefficients as well as the slip length from equilibrium MD simulations.
Under nanometric confinement, we are able to probe the transition to overdamped
sound modes at long wavelengths.

\section{Theory}

\subsection{Dimensionality reduction}
\label{sec:theory_hans}

Continuum formulations of thin-film flows typically employ an average over the
gap height to reduce the dimensionality of the problem.
We have shown in a previous work \citep{holey2022_heightaveraged}, how this can
be formally achieved without making \emph{a priori} assumptions about the
constitutive behavior, which we briefly recap here.
We average the hydrodynamic balance laws over the gap height
\begin{equation}\label{eq:avg_balance}
    \frac{1}{h}\int_{h_1(x,y,t)}^{h_2(x,y,t)} \frac{\partial \mathbf{q}}
     {\partial t}\,\mathrm{d}z =
    -\frac{1}{h}\int_{h_1(x,y,t)}^{h_2(x,y,t)} \left(
    \frac{\partial \mathbf{f}_x}{\partial x} +
    \frac{\partial \mathbf{f}_y}{\partial y} +
    \frac{\partial \mathbf{f}_z}{\partial z} \right)\mathrm{d}z,
\end{equation}
where $\mathbf{q} \equiv \mathbf{q}(\vec{r},t)$ denotes the vector of conserved
densities, and $\mathbf{f}_i \equiv \mathbf{f}_i(\vec{r},t)$ are the
corresponding fluxes in Cartesian direction $i$.
Note that here and in the following, bold symbols (e.g. $\mathbf{q}$, $\mathbf
{f}$) indicate vectors of arbitrary length representing a collection of state
variables (or derived quantities) while arrows (e.q. $\vec{r}$, $\vec
{k}$) indicate Cartesian 3-vectors.
Formally, this integration can be performed for channels with surfaces moving
lateral to each other and having surface topography, such as in lubrication.
Hence, the integration limits $h_1(x,y,t)$ and $h_2(x,y,t)$ depend on the
lateral Cartesian coordinates and on time. 
We denote the gap height with $h \equiv h(x,y,t) = h_2(x,y,t) - h_1(x,y,t)$.

For the integral on the l.\,h.\,s. of Eq.~\eqref{eq:avg_balance} and the first
two terms on the r.\,h.\,s. Leibniz rule for differentiation under the integral
sign applies, and after a few steps~\citep{holey2022_heightaveraged} one
arrives at a dimensionality-reduced form of the balance equations
\begin{equation}\label{eq:avg_balance_2d}
\frac{\partial \bar{\mathbf{q}}}{\partial t}  = - \frac{\partial \bar{\mathbf
 {f}}_x}{\partial x} - \frac{\partial \bar{\mathbf{f}}_y}{\partial y} - \mathbf
 {s},
\end{equation}
where overbars denote height-averages  $\bar{\phi} = \frac{1}{h}\int_{h_1}^
{h_2}\phi\,\mathrm{d}z$, and $\mathbf{s}$ acts as a source term.
Due to the structure of Eq.~\eqref{eq:avg_balance_2d}, where dominant diffusive
fluxes (e.\,g. shear stresses) are lumped into the source term, explicit
numerical schemes for hyperbolic balance equations with source terms, have been
proven successful to solve lubrication problems \cite
{holey2022_heightaveraged}.

In the most general case, the source term contains flux boundary conditions in
the averaging direction, as well as terms which depend on the topography and
movement of the upper and lower wall.
Here, considering only flat channels without shearing, the source term
simplifies to
\begin{equation}\label{eq:src}
\bm{s} = \frac{\mathbf{f}_z\rvert_{z=h_2} - \mathbf{f}_z\rvert_{z=h_1}}{h}.
\end{equation}
The source term vanishes when there is no flux across the boundary, i.\,e. for
impenetrable, perfectly insulating, and slippery walls, where Eq.~\eqref
{eq:avg_balance_2d} describes a two-dimensional fluid.
For nonzero source terms, additional dissipation relative to a laboratory system
is added.
In the following, we focus on systems with impermeable walls under isothermal
conditions.
In order to solve Eq.~\eqref{eq:avg_balance_2d}, we need explicit expressions of
the relevant fluxes as a function of the conserved variables, $\mathbf{f}_i
(\mathbf{q})$ (or their averaged versions), i.\,e. constitutive relations.

\subsection{Hydrodynamic correlations in confined systems}
In the following, we focus on isothermal conditions where the density vector
$\mathbf{q} = (\rho(\vec{r},t),\,\vec{j}(\vec{r},t))^\top$ does not contain the
energy density.
%
  correlation functions of density and longitudinal momentum.
For bulk fluids, this special case is recovered by setting $\gamma=1$, which
renders adiabatic ($c_\mathrm{s}$) and isothermal sound speed ($c_\mathrm
{T}$) equal.
Furthermore, we neglect nonlinear convective terms, which is justified by the
thin-film assumption \citep{karniadakis2005_microflows}.
The flux in Cartesian direction $i\in[x,y]$ and the source term are then given
by
\begin{align}
    \bar{\mathbf{f}}_i = 
    \begin{pmatrix} 
        \bar{j_i} \\ 
        \delta_{xi}\bar{p} - \bar{\tau}_{xi} \\ 
        \delta_{yi}\bar{p} - \bar{\tau}_{yi} 
    \end{pmatrix}, \;
    \mathbf{s} =  \frac{1}{h}
    \begin{pmatrix} 0 \\ 
        \tau_{xz}|_{z=h_2} - \tau_{xz}|_{z=h_1} \\ 
        \tau_{yz}|_{z=h_2} - \tau_{yz}|_{z=h_1} 
    \end{pmatrix},
\end{align}
respectively, where $\delta_{ij}$ is the Kronecker symbol.
Here, $\tau_{ij}$ denotes the components of the viscous stress tensor, which for
a Newtonian fluid in three dimensions reads
\begin{equation}\label{eq:stress_newton}
    \underline{\tau} = \eta \left(\nabla \vec{u} + (\nabla \vec
     {u})^\top \right)+ \left(\zeta -2\eta/3\right) \, \left(\nabla \cdot \vec
     {u}\right)\,\underline{1},
\end{equation}
where $\eta$ and $\zeta$ are the coefficients of shear and bulk viscosity
respectively, $\vec{u}$ is the velocity field, and $\underline{1}$ is the
$3\times3$ unit matrix.
The pressure $p$ is given by a barotropic equation of state (EOS) $p
(\rho)=c_\mathrm{T}^2\rho$, which was chosen to retain the linearity of the
problem.

\begin{figure}[ht]
    \includegraphics[width=\columnwidth]{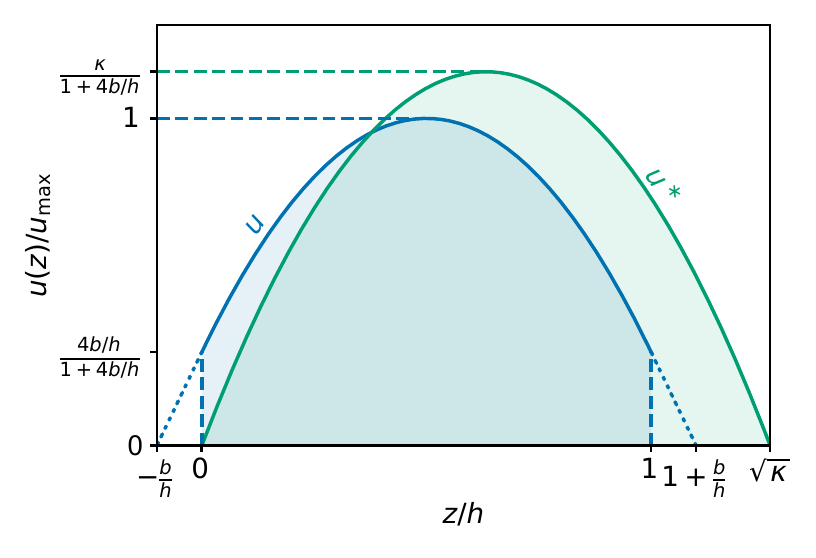}
    \caption{Wall slip leads to an effective gap height $h_\mathrm{eff} = h\sqrt
     {\kappa}$. The parameter $\kappa$ is found by equating the $z$-average of
     the original velocity profile $u(z)$ with that of $u^\ast(z)$, which is
     $u$ shifted to yield zero slip.}
    \label{fig:slip_schematic}
\end{figure}

In nanoscale geometries, deviations from the no-slip boundary conditions become
relevant~\citep{baudry2001_experimental, cheng2002_fluid,
zhu2001_ratedependent}.
Therefore, we consider Navier slip boundary conditions \citep
{navier1823memoire} with a uniform slip length $b$ both at the top and bottom
surface.
The slip length is the virtual distance from the fluid-wall interface at which
the fluid velocity reaches the wall velocity, if linearly extrapolated.
With the assumption that density does not vary across the gap, we obtain the
height-averaged fluxes
\begin{align}
    \bar{\mathbf{f}}_x = 
    \begin{pmatrix} 
        \bar{j_x} \\ c_\mathrm{T}^2\bar{\rho} - \nu_\mathrm{L}\partial_x \bar
         {j}_x \\ 
        - \nu (\partial_x \bar{j}_y + \partial_y \bar{j}_x) 
    \end{pmatrix}, \;
    \bar{\mathbf{f}}_y =
    \begin{pmatrix} 
        \bar{j_y} \\ 
        - \nu (\partial_x \bar{j}_y + \partial_y \bar{j}_x) \\ c_\mathrm
           {T}^2\bar{\rho} - \nu_\mathrm{L} \partial_y \bar{j}_y
    \end{pmatrix},
\end{align}
and the source term
\begin{equation}
    \mathbf{s} =  
    \frac{12\nu}{h^2\kappa} \begin{pmatrix} 0 \\ \bar{j_x} \\ \bar{j_y} \end
     {pmatrix}.
\end{equation}
Here, $\kappa$ renormalizes the actual gap height $h$ for a system with slip to
an effective gap height $h_\mathrm{eff}=h\sqrt{\kappa}$ for an equivalent
system without slip.
The expression for $\kappa$ can be obtained by shifting a parabolic Poiseuille
velocity profile $u(z)$ to no slip boundary conditions while maintaining the
same average flux as in the slip-profile, i.\,e.
\begin{equation}\label{eq:equal_flux}
    \frac{1}{h}\int_0^h u(z) \mathrm{d}z = \frac{1}{h_\mathrm{eff}}\int_0^
     {h_\mathrm{eff}} u^\ast(z) \mathrm{d}z,
\end{equation}
where $ u^\ast(z)$ denotes the shifted profile.
A brief derivation of $\kappa$ for different slip lengths at the top and bottom
wall is given in Appendix~\ref{sec:appendix-kappa}.
In the scope of this work, we only deal with the case of identical slip length
$b$ at both walls, for which we obtain
\begin{equation}
    \kappa=1+6b/h,
    \label{eq:slip-kappa}
\end{equation}
and which is illustrated in Fig.~\ref{fig:slip_schematic}.

\begin{figure}[ht]
    \centering
    \includegraphics[width=\columnwidth]{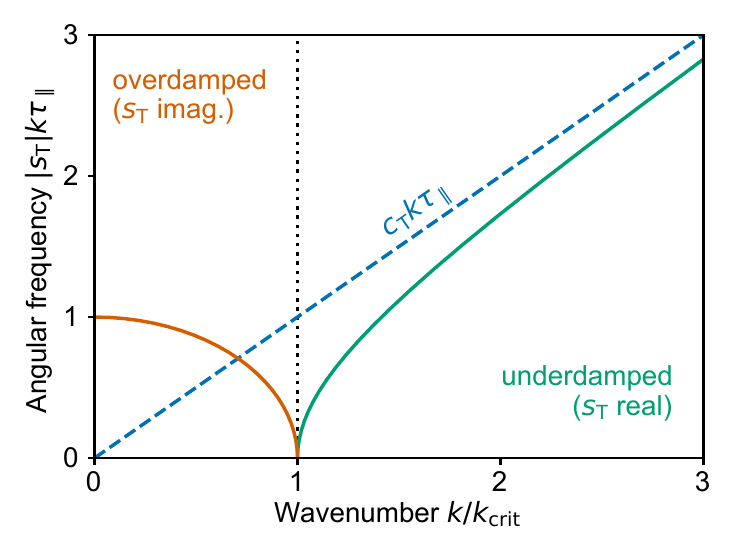}
     \caption{Dispersion relation for a height-averaged continuum formulation of
     confined fluids. 
     The solid line describes the angular frequency using the
     magnitude of the phase velocity, given by Eq.~\eqref{eq:speed_dispersion}. 
     The dashed line describes the bulk reference.
     The wavenumber is normalized by $k_\mathrm{crit} = 6 \nu / h^2 \kappa c_\mathrm{T}$ 
     and the angular frequency is normalized by the characteristic relaxation time
     of longitudinal modes in the underdamped limit 
     $\lim_{k \to k_\mathrm{crit}}\tau_\parallel = h^2 \kappa/ 6 \nu$.
     The critical transition from underdamped to overdamped dynamics occurs
     with diverging group velocity.}
    \label{fig:sound_speed_vs_lambda}
\end{figure}

To arrive at a general solution to Eq.~\eqref{eq:avg_balance_2d}, we express the
densities of conserved variables as a series of normal modes $\mathbf{q}(\vec
{r}, t) = \tilde{\mathbf{q}}(\vec{k},t)e^{\iu\vec{k}\cdot\vec{r}}$, where $\vec
{k}$ and $\vec{r}$ are two-dimensional vectors in the plane of the confined
region.
Note, that for convenience, we keep the notation introduced above, but all
vector dimensions are reduced by one due to the average.
Since all field variables are averaged over the gap height, we drop overbars
from now on for the sake of brevity.
We obtain an ordinary differential equation for the Fourier coefficients
\begin{equation}\label{eq:isoT_2D_ode}
    \frac{\mathrm{d}\tilde{\mathbf{q}}(k,t)}{\mathrm{d}t} = \mathbf
     {H}\cdot\tilde{\mathbf{q}}(k,t),
\end{equation}
with the hydrodynamic matrix
\begin{equation}\label{eq:hydro_matrix}
    \mathbf{H} = - 
    \begin{bmatrix} 0 & \iu k & 0 \\
        \iu c_\mathrm{T}^2k & \nu_\mathrm{L}k^2 + 12\nu/(h^2\kappa) & 0 \\ 0 &
         0 & \nu k^2 + 12\nu/(h^2\kappa)
    \end{bmatrix},
\end{equation}
The hydrodynamic matrix $\mathbf{H}$ is diagonalized with eigenvalues
\begin{align}\label{eq:shear_ev}
    \mu_\perp = -\nu k^2 - \frac{12\nu}{h^2\kappa},
\end{align}
corresponding to transverse modes, and
\begin{align}\label{eq:longitudinal_ev}
    \mu_\parallel = -\frac{\nu_\mathrm{L}}{2}k^2 - \frac{6\nu}
     {h^2\kappa} \pm \iu s_\mathrm{T} k,
\end{align}
corresponding to longitudinal modes.
Here, the isothermal speed of sound $s_\mathrm{T}(k) = \sqrt{c_\mathrm{T}^2 -
(\tau_\parallel k)^{-2}}$ follows a confinement-induced dispersion relation
with $\tau_\parallel=(\operatorname{Re}\,\mu_\parallel)^{-1}$ which for small
wavenumbers ($hk\ll 1$) reads
\begin{equation}\label{eq:speed_dispersion} 
s_\mathrm{T}(k) = \sqrt{c_\mathrm{T}^2 - \frac{36\nu^2}{h^4\kappa^2 k^2}}.
\end{equation}
Surprisingly, the dispersion takes effect at large wavelengths $\lambda = 2\pi /
k$, where the second term in the discriminant becomes important, and the speed
of sound deviates from its bulk counterpart given by the EOS.
As long as $s_\mathrm{T}$ is a real number, longitudinal modes show underdamped
oscillations.
However, there is a transition from underdamped to overdamped behavior at a
critical wavelength
\begin{equation}
    \lambda_\mathrm{crit} = \frac{2\pi}{k_\mathrm{crit}} = \pi h^2 \kappa
     c_\mathrm{T} /3\nu,
    \label{eq:lambda-crit}
\end{equation}
where $s_\mathrm{T}$ becomes imaginary and the eigenvalues for longitudinal
modes $\mu_\parallel$ become real.
In Fig.~\ref{fig:sound_speed_vs_lambda}, we illustrate the wavelength dependence
of the effective speed of sound.
Note, that we plot absolute values for $s_\mathrm{T}$ normalized by
$k \tau_\parallel$ as a function of the wavenumber normalized by $k_\mathrm
{crit}$.
Similar expressions can be derived for axisymmetric flow through circular
channels with radius $R$, where the only two eigenvalues are given by
\begin{equation}
    \mu^{\mathrm{1D}}_\parallel = -\frac{4\nu}{R^2\kappa} \pm \iu \sqrt
     {c_\mathrm{T}^2-\frac{16\nu^2}{R^4\kappa^2 k^2}}k.
\end{equation}
with $\kappa=1+4b/R$.
The definition of derived quantities such as $\lambda_\mathrm{crit}$ change
accordingly.

Consequently, modes with wavelengths larger than the critical wavelength
$\lambda_\mathrm{crit}$ cannot propagate through the channel and are dissipated
within a finite relaxation time.
This is rather counterintuitive, since in bulk hydrodynamics
wavelength-independent transport coefficients are assumed at sufficiently long
wavelengths.

Assuming arbitrary initial conditions $\tilde{\mathbf{q}}(k, 0) = (\tilde{\rho}
(k, 0), \tilde{j}_\parallel(k, 0), \tilde{j}_\perp(k, 0))^\top$ (with the
limitation that perturbations out of equilibrium are small in order to preserve
linearity), we solve Eq.~\eqref{eq:isoT_2D_ode} for the time evolution of the
real part of the conserved variables,
\begin{subequations}
\label{eq:acf_slab_underdamped}
\begin{align}
\frac{\tilde{\rho}(k,t)}{\tilde{\rho}(k, 0)} & = e^{-t/\tau_\parallel} \left
 [\cos(s_\mathrm{T} k t) + \frac{1}{s_\mathrm{T} k \tau_\parallel} \sin
 (s_\mathrm{T} k t)\right] \label{eq:acf_dens} \\
\frac{\tilde{j}_\parallel(k,t)}{\tilde{j}_\parallel(k,0)} &=  e^
 {-t/\tau_\parallel} \left[\cos(s_\mathrm{T} k t) - \frac{1}{s_\mathrm
 {T}k\tau_\parallel}\sin(s_\mathrm{T} k t)\right] \label{eq:acf_long}\\
\frac{\tilde{j}_\perp(k, t)}{\tilde{j}_\perp(k, 0)} &=  e^
 {-t/\tau_\perp}, \label{eq:acf_trans}
\end{align}
\end{subequations}
with characteristic relaxation times 
\begin{subequations}\label{eq:relax}
\begin{align}
\tau_\perp(k) &= (\nu k^2 + 12\nu/h^2\kappa)^{-1}, \label{eq:relax_trans}\\
\tau_\parallel (k) &= (\nu_\mathrm{L} k^2/2 + 6\nu/h^2\kappa)^{-1}. \label
 {eq:relax_long}
\end{align}
\end{subequations}
Note, that these expressions remain valid when $s_\mathrm{T}$ becomes purely
imaginary and the trigonometric functions turn into their hyperbolic
counterparts.
We give the solution of the imaginary parts explicitly in Appendix~\ref
{sec:appendix_imag} for completeness.

\begin{figure}[!ht]
    \includegraphics[width=\columnwidth]{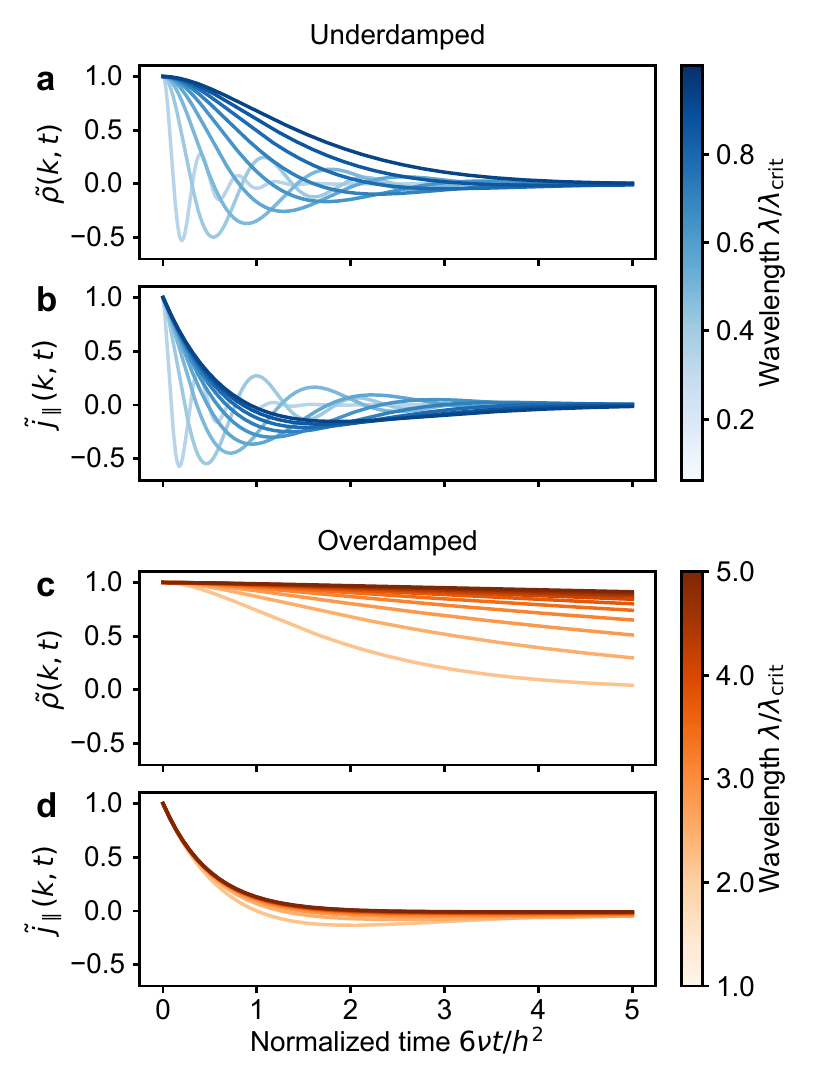}
    \caption{Fourier coefficients for density and longitudinal momentum in the
     underdamped and overdamped case, respectively.}
    \label{fig:slab_theory}
\end{figure}

Figures~\ref{fig:slab_theory}a and b show Eq.~\eqref{eq:acf_dens} and \eqref
{eq:acf_long}, respectively, as well as their dependence on wavelength in the
case of underdamped oscillations ($\lambda<\lambda_\mathrm{crit}$).
Since $\lambda_\mathrm{crit}\propto c_\mathrm{T} h^2 / \nu$, and $c_\mathrm
{T}/\nu \sim \mathcal{O}(1)$ for most dense fluids, the critical wavelength is
on the order of the magnitude of the squared gap height.
In the overdamped case ($\lambda>\lambda_\mathrm{crit}$), $s_\mathrm{T}$ is an
imaginary number, and therefore, the behavior of density and longitudinal
momentum modes changes fundamentally as shown in Fig.~\ref
{fig:slab_theory}c--d.
We observe that density perturbations do not decay, i.\,e. their relaxation time
diverges.
Furthermore, one can show that the decay rate scales with $k^2$, similar to bulk
fluids (see Appendix~\ref{sec:appendix_overdamped}).
However, in contrast to the bulk, sound relaxation times converge to a finite
value, identical to that of transverse modes.

We want to highlight, that the functional form of Eq.~\eqref
{eq:acf_slab_underdamped} is equivalent to the autocorrelation function of
equilibrium fluctuations in the bulk, but with different characteristic time
scales, as described above.
In particular, relaxation times of longitudinal and transverse modes become
wavelength-independent in the long wavelength limit, and a diverging group
velocity leads to a transition to overdamped sound relaxation.
We explicitly probed this transition using equilibrium molecular dynamics
simulations of confined fluids.
Therefore, we write generic correlation functions of the form
\begin{subequations}
\label{eq:acf_generic}
\begin{align} C_\perp(k,t) &= e^{-\frac{t}{\tau_\perp}}, \\
    \label{eq:acf_generic_long} C_\parallel(k,t) &= e^{-\frac{t}
     {\tau_\parallel}} \left[\cos(\omega t) - \beta  \sin(\omega t)\right], \\
    \label{eq:acf_generic_density} C_\rho(k,t) &= \frac{\gamma -1}
     {\gamma}e^\frac{t}{\tau_\mathrm{T}} + \frac{1}{\gamma} e^{-\frac{t}
     {\tau_\parallel}} \left[\cos(\omega t) + \beta  \sin(\omega t)\right],
\end{align}
\end{subequations}
where $\beta = 1 /\omega \tau_\parallel$, and use frequencies and decay rates as
fit parameters to the autocorrelation functions obtained from MD. 
Note that for bulk systems, since $\beta\sim\mathcal{O}(k)$, the sine-term is
often neglected, but some authors \citep
{berne1971_time, schoen1986_computation, porcheron2002_propagating} explicitly
include it.
As highlighted in the beginning of this section, our derivation is for
non-fluctuating isothermal conditions with $\gamma=1$.
However, the numerical tests with MD naturally contain thermal fluctuations at
macroscopic constant temperature, which is why we included the thermal part in
Eq.~\eqref{eq:acf_generic_density}.
An overview of the theoretical expressions for $\omega$ and $\tau$ for different
systems and conditions is given in Tab.~\ref{tab:time_const}.

\begin{table}
    \caption{Theoretical expressions for the characteristic time scales for wave
     propagation and decay in bulk and confined systems.}
    \label{tab:time_const}
    \begin{ruledtabular}
    \begin{tabular}{lcccc}
        & $1/\tau_\mathrm{T}$ & $1 / \tau_\perp$ & $1 / \tau_\parallel$ &
          $ \omega$ \\
        \colrule
        & & & & \\
        bulk & $D_\mathrm{T}k^2$ & $\nu k^2$ & $\Gamma k^2$ & $c_\mathrm
        {s} k$ \\ bulk ($\gamma=1$) & -- & $\nu k^2$ & $\frac{\nu_\mathrm
        {L}k^2}{2}$ & $c_\mathrm{T} k$ \\ confined ($\gamma=1$) & -- & $ \nu
        k^2 + \frac{12\nu}{h^2\kappa} $ & $\frac{\nu_\mathrm{L}k^2}{2} + \frac
        {6\nu}{h^2\kappa}$ & $s_\mathrm{T}(k) k$ \\
    \end{tabular}
    \end{ruledtabular}
\end{table}

\section{Molecular Dynamics Simulations}

In the previous section, we derived expressions for the Fourier coefficients of
conserved variables in confined fluids in the hydrodynamic limit.
To scrutinize our predictions, we performed a brute force test of our findings
using molecular dynamics (MD) simulations of a simple fluid, confined in a
nanometer-sized channel.
All MD simulations were carried out with LAMMPS \citep{thompson2022_lammps}, and
we have used a supercritical Lennard-Jones fluid at the state point
$T=2.0\,\epsilon/k_\mathrm{B}$ and $\rho=0.452\,\sigma^{-3}$ for all results
shown in this paper.
Yet, our findings are not limited to the supercritical state, which we showed in
a related work \citep{holey_confinementinduced}.
Hence, the interaction potential~\citep{muser2023_interatomic} between fluid
atoms is given by
\begin{equation} U(r_{ij}) = 4 \epsilon\left[\left(\sigma/r_{ij}\right)^
 {12} - \left(\sigma/r_{ij}\right)^{6}\right],
\end{equation}
where $r_{ij} = |\vec{r}_i - \vec{r}_j|$ is the distance between particle $i$
and $j$. The interatomic potential is shifted to zero at a cutoff radius of
$r_\mathrm{c}=2.5\sigma$.
All simulations were performed with a time step $\Delta t=0.0025\sqrt
{m\sigma^2/\epsilon}$.

We performed bulk reference simulations in fully periodic boxes, as well as
simulations in confined system, where the periodicity is broken in the
direction normal to the walls, which are modeled as explicit rigid atoms.
To probe the long wavelength limit, we used simulation boxes with high aspect
ratios, i.\,e. where one box length in the direction parallel to the walls is
much larger than both the gap height and the remaining in-plane dimension.
We sampled the trajectories in the microcanonical ensemble for $2\times10^7$
time steps after an initial equilibration of the system in the canonical
ensemble, and recorded every 2000th step for post-processing.

\subsection{Confined setup}
\label{sec:slab_setup}

For the confined fluid simulations, we model the walls as two rigid atomic
layers arranged in an \emph{fcc} lattice with lattice constant
$a=1.2\sigma_\mathrm{f}=1.2\sigma$, which corresponds to a wall density of
$\rho_\mathrm{w} = 2.31 \sigma^{-3}$.
The wall density significantly affects fluid-wall commensurability, hence
slip~\citep{thompson1990_shear}, and the chosen parameter lies within the range
of similar studies \citep{thompson1990_shear, jabbarzadeh1999_wall,
priezjev2004_molecular}.
The $[1\,1\,\bar{2}]$ and $[\bar{1}\,1\,0]$ directions are taken in $x$- and
$y$-direction, respectively, such that the closest-packed $\{ 111\}$-surface is
in contact with the fluid. In $x$- and $y$-direction, we keep periodic boundary
conditions, hence the lateral box sizes are always chosen to be multiples of
$3a/\sqrt{2}$ and $a/\sqrt{2}$ respectively.
Fluid-wall interactions are governed by a Lennard-Jones potential.
We use the Lorentz mixing rule for the length scales $\sigma_\mathrm{wf}=
(\sigma_\mathrm{w} + \sigma_\mathrm{f}) / 2$, with $\sigma_\mathrm
{w}=0.75\sigma$.
We model different wetting behavior at the interface by scaling the fluid-wall
interaction with respect to the fluid's interaction energy, i.\,e.
$\epsilon_\mathrm{wf}=\alpha \epsilon_\mathrm{f}$, where the parameter $\alpha$
can be related to the contact angle \citep{yamaguchi2019_interpretation}.

\begin{figure}[h]
    \includegraphics[width=\columnwidth]{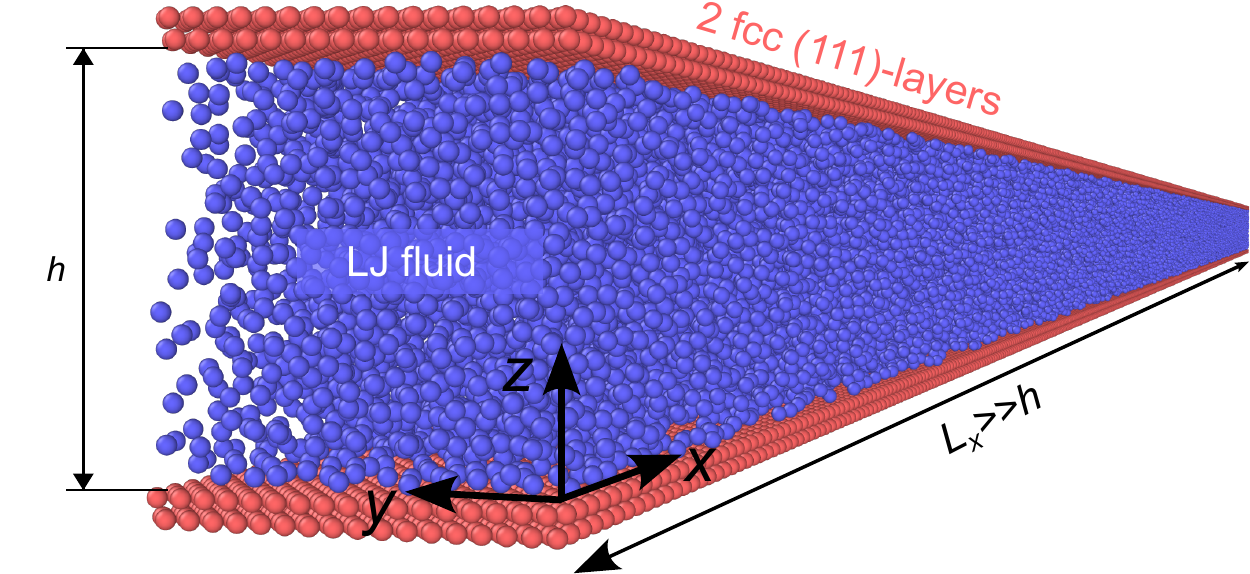}
    \caption{Simulation setup for molecular dynamics simulations of confined
     fluids. One lateral dimension is much larger than the gap height.}
    \label{fig:md_setup}
\end{figure}

Input parameters for the confined system simulations are gap height, lateral box
sizes, and number density, which defines the number of atoms $N$ in a
homogeneous system.
Note, that the inner wall layers are placed at a distance $h+\sigma$ apart to
account for a thin depletion zone at the interface.
In the case of pronounced layering effects, fixing the gap height can lead to
deviations from the target density.
Yet, we stick to this pragmatic approach, which is particularly useful when
comparing atomistic results with continuum predictions.
For the gap heights and wetting properties covered within this study, we did not
observe strong deviations from the target density in the center of the fluid
film.
The confined MD setup is illustrated in Fig.~\ref{fig:md_setup}

\subsection{Autocorrelation functions from equilibrium fluctuations}

For both bulk and confined systems, we compute the autocorrelation functions for
the Fourier coefficients of mass and momentum density.
We compute the average in Eq.~\eqref{eq:acf_transverse}--\eqref
{eq:acf_density} from a single time series of the dynamical variables, which is
equivalent to the convolution of that time series with itself.
We leverage the computational efficiency of the Fast Fourier Transform (FFT) and
use the Wiener-Khinchin theorem to compute the integral \citep
{chatfield2003_analysis}.
We chose $\vec{k}=(k_n,0,0)^\top$ in the direction of the longest dimension of
our simulation box, where $k_n$ is a discrete wavevector corresponding to waves
that fit into this periodic dimension, i.\,e. $k_n=2\pi
n/L_x\;n=1,\,2,\,3,\ldots$.
Hence, momentum autocorrelation functions are calculated for the transverse
($\vec{j} \perp \vec{k}$) and longitudinal ($\vec{j} \parallel \vec
{k}$) direction.

To rule out possible size or shape effects in this setup, we varied the lateral
box sizes independently and compared transverse momentum autocorrelation
functions computed from either of the lateral velocity components, which led to
indistinguishable results.
Hence, computing transport coefficients from correlation functions of collective
variables in boxes with high aspect ratio seems to be no issue, in contrast to
e.\,g. the calculation of self-diffusion coefficients~\citep
{yeh2004_systemsize, kikugawa2015_effect}.

\subsection{Non-equilibrium simulations of slip}
\label{sec:nemd}

We performed non-equilibrium MD simulations with a similar setup as described in
Sec.~\ref{sec:slab_setup}.
Instead of sampling the equilibrium state, we sheared the walls at a constant
shear rate, to obtain reference values for the slip length.
Therefore, we used smaller box sizes in the shearing direction ($L_x =
29.40\sigma$) resulting in 2818 fluid atoms.
We sheared the upper and lower wall at constant velocity $\pm u_0/2$,
respectively, and sampled the Couette profile after an initial startup period
until a total sliding distance of $100L_x$ is reached.

Shearing at a fixed gap height with rigid wall atoms may lead to shear
localization \citep{gattinoni2014_boundarycontrolled} or diverging slip
lengths \citep{martini2008_slip}.
The choice of our simplified setup is mainly to ensure better comparison with
equilibrium simulations, where these effects do not play a role.
Here, we did not observe such phenomena at the densities and shear rates
considered.
Furthermore, the maximum applied shear rate $u_0/h$ in our reference simulations
is lower than $0.01\,\sqrt{\epsilon/m\sigma^2}$, where the shear rate
dependence is expected to be low, even for strong fluid wall interaction \citep
{priezjev2007_ratedependent}.

\subsection{Green-Kubo simulations}
\label{sec:gk}

We performed bulk equilibrium simulations in cubic boxes with 1000 atoms to
obtain reference viscosities using the Green-Kubo approach.
The GK integrals for the shear and bulk viscosity read
\begin{subequations}
\begin{align}
\eta &= \frac{V}{k_\mathrm{B} T}\int_0^\infty \langle \tau_{ij} (t) \tau_{ij}
 (0)\rangle \mathrm{d}t ,\label{eq:gk} \\
\zeta &= \frac{V}{k_\mathrm{B} T}\int_0^\infty \langle \delta p (t) \delta p
 (0)\rangle \mathrm{d}t , \label{eq:gk_bulk}
\end{align}
\end{subequations}
respectively, where $\tau_{ij}$ are the components of the deviatoric stress
tensor and $\delta p (t) = \mathrm{tr}\left[\underline{\sigma}
(t) \right]/3 - \langle p \rangle$ with stress tensor $\underline{\sigma}
(t)$ and the equilibrium average of the hydrostatic pressure $\langle
p \rangle$. 
The value of the integrals in Eq.~\eqref{eq:gk} and \eqref{eq:gk_bulk} converges
after approximately $500 \Delta t$, for correlation functions computed from a
trajectory with $4\times 10^5$ time steps.
The viscosities have been computed from the integrals of 20 equivalent replica
simulations.

\section{Results}
\subsection{Autocorrelation functions}\label{sec:results_acf}

We first computed the autocorrelation functions from a bulk reference simulation
without walls.
The simulation box had dimensions $941.2\times14.7\times14.7\,\sigma^3$, which
corresponds to $92\,001$ atoms.
Confined system simulations were performed in a similar box with $90\,189$ fluid
atoms and $43\,520$ solid atoms in each of the two walls.
In this section, we show simulation results with a wall-fluid interaction
parameter $\alpha=0.75$.
Note, that typical values for $\sigma$ are in the range of a few
{\AA}ngströms, and therefore, our systems have gap heights of a few
nanometers.
We obtained characteristic times for the decay and propagation of equilibrium
fluctuations by a least-square fit to the generic expressions, Eq.~\eqref
{eq:acf_generic}, for discrete wavenumbers.
We used MD data in a time interval of $250\,\sqrt{m\sigma^2/\epsilon}$ for the
fit, and for the confined system all wavelengths considered at this point are
in the underdamped regime.
In all cases, we used rates (e.\,g. $1/\tau_\perp$) and frequencies($\omega$) as
fitting parameters, but plot the inverse, i.\,e. the corresponding
characteristic times.

\begin{figure*}[ht]
    \centering
    \includegraphics[width=\linewidth]{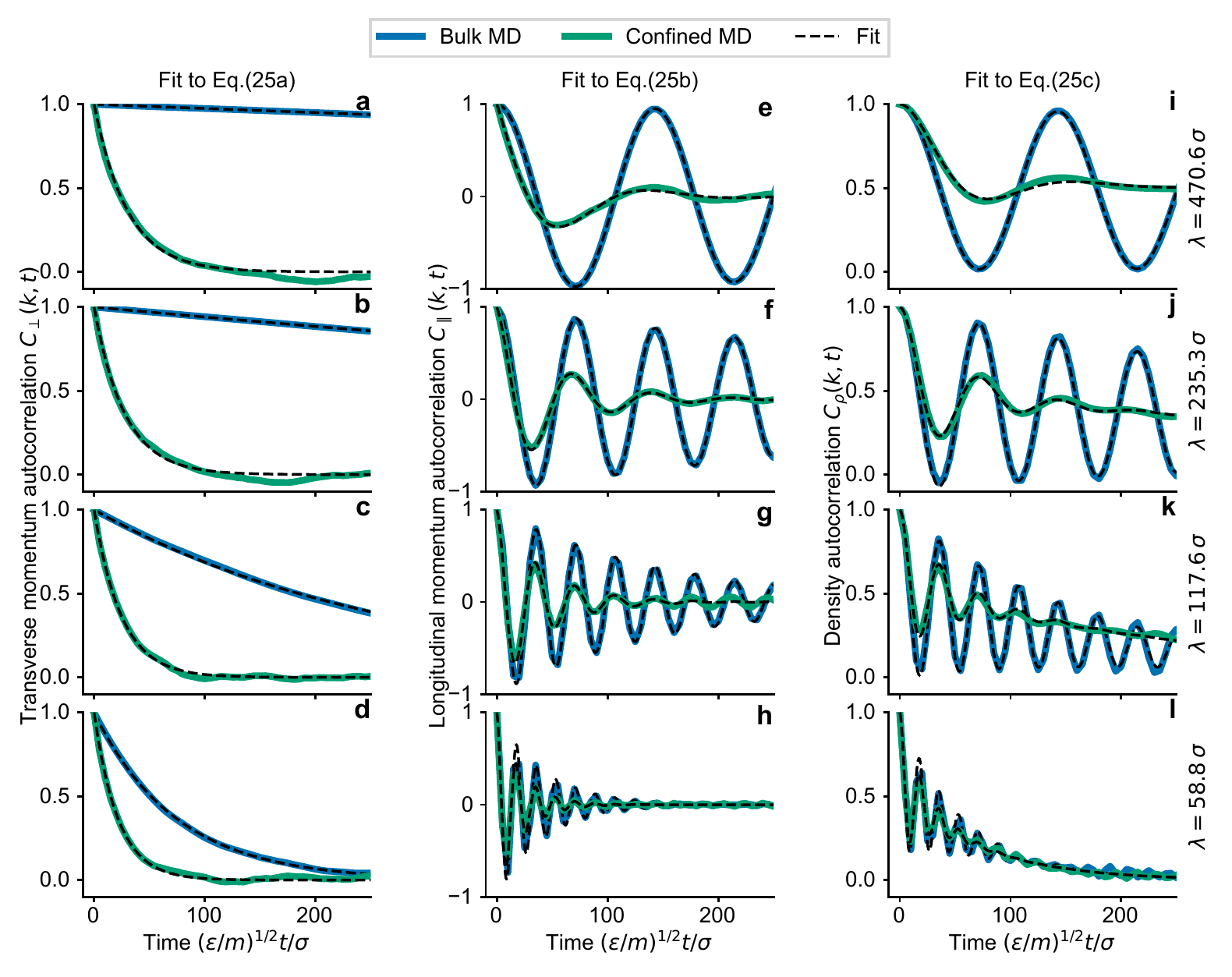}
    \caption{Autocorrelation functions of the real part of density and momentum
     modes. The first column (a-d) corresponds to transverse momentum ($\vec
     {j}\perp\vec{k}$), the second column  (e-h) corresponds to longitudinal
     momentum ($\vec{j}\parallel\vec{k}$), and the third column
     (i-l) corresponds to density correlations. Panels within the same row
     share the same wavelength. Solid lines show results from molecular
     dynamics (MD) calculations, and dashed lines are the corresponding fits.
     For simulations of the confined system, all wavelengths shown here are in
     the underdamped regime ($470.6\sigma < \lambda_\mathrm{crit}$).}
    \label{fig:acf_underdamped}
\end{figure*}

Transverse momentum correlations are independent of the longitudinal modes, and
the kinematic shear viscosity $\nu$ is the corresponding transport
coefficient.
Figures~\ref{fig:acf_underdamped}a--d show the autocorrelation functions of the
real part of $j_y(k, t)$ for four wavelengths (blue solid lines) and the
corresponding fit (black dashed lines), where the shear relaxation time
$\tau_\perp$ increases with wavelength.

From the transverse momentum autocorrelation functions of the confined fluid in
Fig.~\ref{fig:acf_underdamped}a--d it becomes evident, that the wavelength
dependence in the decay rate is lost in the range of the presented
wavelengths.
This is qualitatively in agreement with the prediction in Eq.~\eqref
{eq:acf_slab_underdamped}. 
The relaxation times of transverse momentum fluctuations are much shorter than
in the bulk.

The autocorrelation functions of longitudinal momentum fluctuations obtained
from MD simulations as well as the corresponding fits are shown in Fig.~\ref
{fig:acf_underdamped}e--h.
Different relaxation behavior between the bulk and the confined systems can be
seen in the longitudinal direction as well, with stronger damping in the system
with walls.
For the shorter wavelengths, the positions of the local minima and maxima in the
autocorrelation functions coincide with those of the bulk fluid.
For the largest wavelength in Fig.~\ref{fig:acf_underdamped}e, we observe that
the first minimum is shifted to shorter times compared to the bulk, indicating
a frequency shift with increasing wavelengths.

The density autocorrelation function Eq.~\eqref{eq:acf_density} contains the
previously obtained longitudinal momentum autocorrelation function weighted by
$1/\gamma$.
Therefore, to reduce the number of fitting parameters, we fix the sound
attenuation rate and sound frequency, which leaves only the decay rate of
thermally induced density fluctuations $1/D_Tk^2$ and the heat capacity ratio
$\gamma$ to be determined.
The results are shown in Fig.~\ref{fig:acf_underdamped}i--l, where we make
similar observations of faster relaxation with weaker wavelength-dependence as
in the bulk fluid.

\subsection{Effective time scales under confinement}
\label{sec:results_predict}

The correlations of equilibrium fluctuations in confined fluids clearly deviate
from their bulk counterparts, as shown in the previous section, but Fig.~\ref
{fig:acf_underdamped} highlighted only four selected wavelengths.
In the following, we systematically investigate the transition of characteristic
time scales as wavelength increases.
In particular, we test whether the isothermal, height-averaged theory described
in Sec.~\ref{sec:theory_hans} adequately describes the relaxation times and
frequencies obtained from MD.
Therefore, we computed all quantities appearing in Eq.~\eqref
{eq:relax_long}, \eqref{eq:relax_long}, and \eqref{eq:speed_dispersion} in
separate non-equilibrium (see Sec.~\ref{sec:nemd}) and equilibrium MD
simulations (see Sec.~\ref{sec:gk}), and compare the theoretical predictions
with the corresponding fit parameters from the autocorrelation functions.
Since, thermal effects are not considered here, but are naturally included in
the MD simulations, we obtained additional parameters ($C_\mathrm
{P}$, $C_\mathrm{V}$, $\kappa_\mathrm{T}$) from the NIST thermophysical
database for Argon \cite{lemmon2023_thermophysical}.
A summary of material parameters at the supercritical state point is given in
Tab.~\ref{tab:bulk_fitting_const}.

\begin{table}[ht]
    \caption{Material parameters obtained for a supercritical
     ($T=2.0\,\epsilon/k_\mathrm{B}$, $\rho=0.452\,\sigma^{-3}$) Lennard-Jones
     (LJ) fluid obtained from molecular dynamics (MD) simulations, and from the
     NIST database for supercritical Argon. Derived quantities such as e.\,g.
     the sound attenuation coefficient $\Gamma$ may depend on values from both
     sources. All reported parameters are used for the predictions in
     Figs.~\ref{fig:fit_relax_bulk-vs-slab} and \ref
     {fig:fit_freq_bulk-vs-slab}.}
    \label{tab:bulk_fitting_const}
    \begin{ruledtabular}
    \begin{tabular}{lccc} Name & symbol & value & unit \\
        \colrule Shear viscosity$^\ast$ & $\eta$ & $0.549$ & $\sqrt
         {m\epsilon}/\sigma^2$\\ Bulk viscosity$^\ast$ & $\zeta$ & $0.351$ &
         $\sqrt{m\epsilon}/\sigma^2$\\ Kinematic shear viscosity$^\ast$ &
         $\nu$ & $1.215$ & $\sigma\sqrt{\epsilon/m}$ \\ Kinem. longitudinal
         viscosity$^\ast$ & $\nu_\mathrm{L}$ & $2.396$ & $\sigma\sqrt
         {\epsilon/m}$ \\ Thermal diffusivity$^\dagger$ & $D_\mathrm
         {T}$ & $0.548$ & $\sigma\sqrt{\epsilon/m}$ \\ Sound attenuation
         coefficient$^{\ast, \ddagger}$ & $\Gamma$ & $1.627$ & $\sigma\sqrt
         {\epsilon/m}$ \\ Heat capacity ratio$^\dagger$ & $\gamma$ &
         $2.567$ & -- \\        Isothermal speed of sound$^\dagger$ &
         $c_\mathrm{T}$ & $1.997$ & $\sqrt{\epsilon/m}$ \\ Adiabatic speed of
         sound$^{\dagger,\ddagger}$ & $c_\mathrm{s}$ & $3.198$ & $\sqrt
         {\epsilon/m}$ \\
    \end{tabular}
    \end{ruledtabular}
    \vskip1ex
    \raggedright
    \footnotesize $^\ast$ from MD (GK) \hspace{2ex} $^\dagger$ from MD
     (EOS) \hspace{2ex} $^\ddagger$ from NIST \cite{lemmon2023_thermophysical}
\end{table}

\begin{figure}[!ht]
    \centering
    \includegraphics[width=\columnwidth]{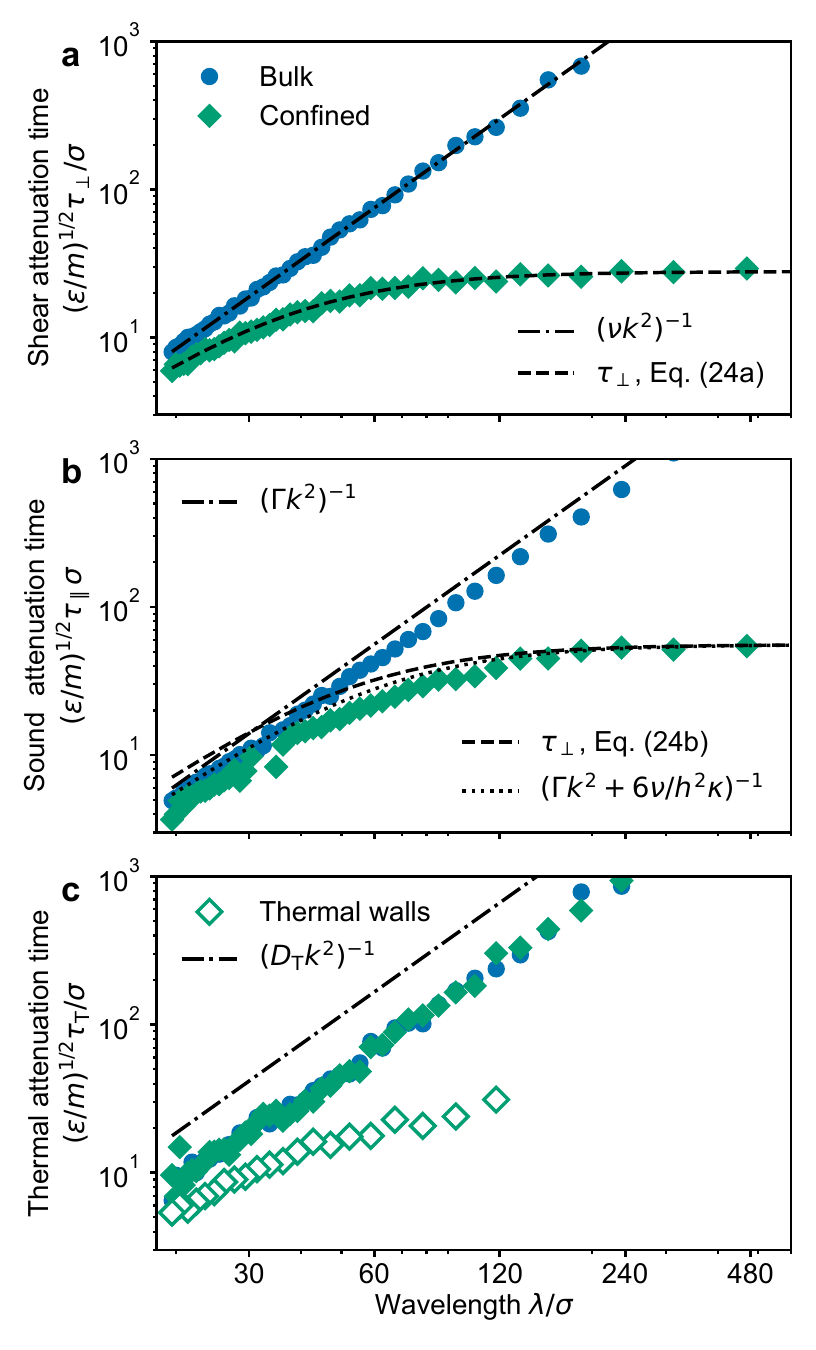}
    \caption{Relaxation times obtained from fits to the autocorrelation
     functions of momentum and density fluctuation as a function of the
     wavelength for bulk and confined systems with $\alpha=0.75$. For all three
     subfigures, simulation results for the bulk system are shown as blue
     discs, and results for the confined system are shown as green diamonds.
     Panels a and b show the theoretical prediction for bulk and confined
     systems as dash-dotted and dashed lines, respectively. In panel b, the
     relaxation times according to the isothermal theory, Eq.~\eqref
     {eq:relax_long}, are modified for short wavelengths by considering thermal
     effects, which is illustrated as a dotted line. Panel c shows thermal
     relaxation times obtained from density autocorrelation functions, which
     have quadratic wavelength scaling for both bulk and confined systems, as
     long as walls are rigid. Open diamonds in panel c show results from
     simulations with thermal walls, which indicate a similar transition as in
     a and b.}
    \label{fig:fit_relax_bulk-vs-slab}
\end{figure}

Figure.~\ref{fig:fit_relax_bulk-vs-slab}a shows the shear attenuation time
$\tau_\perp$ over the wavelength as obtained from both bulk and confined MD
simulations.
The symbols represent an average over the two transverse directions ($j_y(k, t)$
and $j_z(k,t)$, real and imaginary parts respectively) in the bulk, and over
the in-plane transverse direction (from $\operatorname{Re}[j_y(k,t)]$ and
$\operatorname{Im}[j_y(k,t)]$) in the confined system.
We do not investigate correlations of the $j_z$-Fourier coefficients, since we
assumed laminar flow to motivate the height-averaged balance equations in
Sec.~\ref{sec:theory_hans}.

Shear relaxation times in the bulk scale with the square of the wavelength as a
consequence of momentum conservation.
The dash-dotted line in Fig.~\ref{fig:fit_relax_bulk-vs-slab}a illustrates the
prediction for the bulk fluid with kinematic viscosity as the constant of
proportionality.
As expected, the shear relaxation time in confined systems deviates from the
bulk with increasing wavelength, and converges to a constant value of
approximately $30\sqrt{m\sigma^2/\epsilon}$ for wavelengths larger than
$120\sigma$.
The prediction for the confined fluid based on Eq.~\eqref{eq:relax_trans} is
shown as a dashed line, and is in excellent agreement with the MD data.

Similar behavior can be observed for the sound attenuation time $\tau_\parallel$
in Fig.~\ref{fig:fit_relax_bulk-vs-slab}b.
Here, symbols for both bulk and confined configuration correspond to an average
over the fitted relaxation times from the real and imaginary part of the
longitudinal momentum correlations.
The bulk relaxation times clearly follow a quadratic scaling relation, and the
dash-dotted line illustrates the prediction with shear-attenuation coefficient
$\Gamma$ given in Tab.~\ref{tab:bulk_fitting_const}.
For the confined fluid, the transition to wavelength-independent relaxation
times is similar to the in-plane shear relaxation.
Here, the characteristic time for the decay of sound modes converges to a value
approximately twice as large as the shear relaxation time in the
long-wavelength limit, and to the same quadratic scaling as in the bulk in the
limit of short wavelengths.
The prediction based on the isothermal theory, Eq.~\eqref{eq:relax_long}, is
shown as a dashed line.
In the long wavelength limit, relaxation times from MD and the predictions
agree, but the isothermal theory underestimates sound attenuation at shorter
wavelengths.
Therefore, we plot a modified prediction as a dotted line, which replaces the
first term of Eq.~\eqref{eq:relax_long} with the non-isothermal sound
attenuation from the bulk theory, which describes the MD data slightly better.

For completeness, we show the relaxation behavior associated with the Rayleigh
process, i.\,e. thermal relaxation times obtained from the density
autocorrelation functions, in Fig.~\ref{fig:fit_relax_bulk-vs-slab}c.
The dash-dotted line describes again the theoretical expectation based on the
bulk theory, where the thermal diffusivity $D_\mathrm{T}$ is entirely based on
data from NIST.
Thermal relaxation times from bulk and confined MD simulations both scale with
the square of the wavelength and are indistinguishable from one another in the
case of rigid walls.
We show additional data obtained from simulations in shorter boxes with
vibrating wall atoms with open symbols, which indicates a similar transition to
wavelength-independent relaxation times if energy transport through the walls
is allowed.

\begin{figure}[!ht]
    \centering
    \includegraphics[width=\columnwidth]{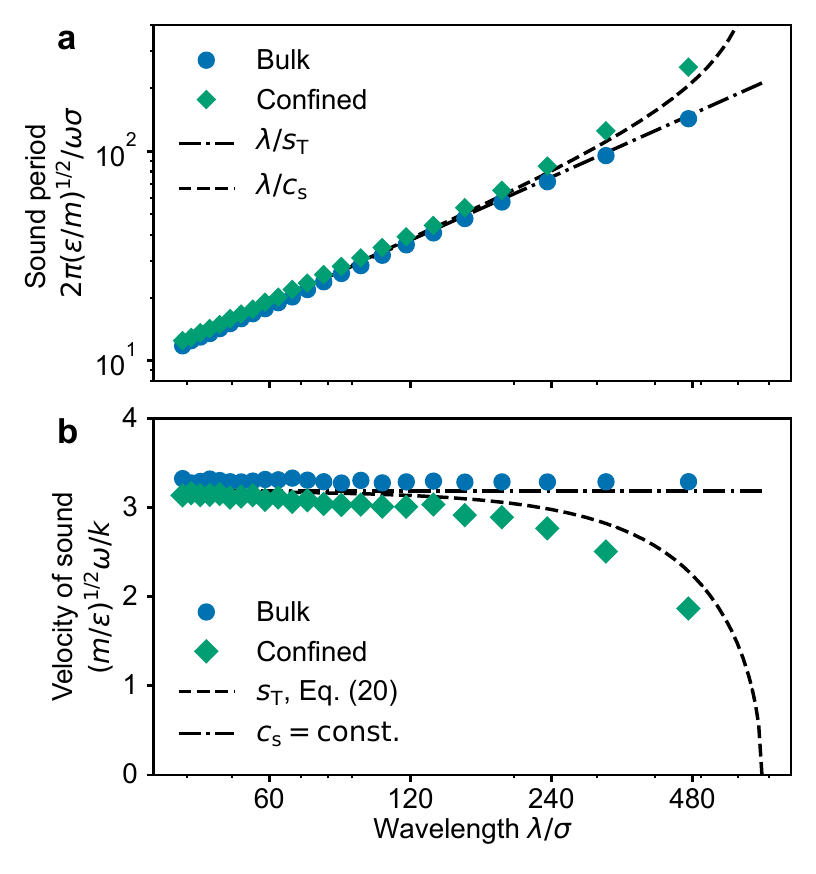}
    \caption{Effective sound period and velocity of sound as obtained from
     longitudinal momentum correlations for bulk and confined systems from
     simulations with $\alpha=1.5$. In panel a, we observe an overall shift to
     longer sound periods for the confined fluids, which increases with
     wavelength, and panel b compares the sound velocities directly. For the
     confined system, the prediction based on the dispersion relation
     Eq.~\eqref{eq:speed_dispersion} adequately describes the deviation from
     the bulk. The critical wavelength marks the transition from underdamped to
     overdamped behavior at $\lambda_\mathrm{crit}=675\,\sigma.$}
    \label{fig:fit_freq_bulk-vs-slab}
\end{figure}

Oscillatory behavior of density and longitudinal momentum correlation encodes
sound propagating properties of the considered fluid systems.
In Fig.~\ref{fig:fit_freq_bulk-vs-slab}a, we plot the sound period
$T=2\pi/\omega$ over the wavelength for both bulk and confined fluids, obtained
from fits to the theoretical expression for the momentum autocorrelation
function.
For illustrative purposes, we chose stronger wall-fluid interactions with
$\alpha=1.5$, as compared to $\alpha=0.75$ in Fig.~\ref
{fig:acf_underdamped} and \ref{fig:fit_relax_bulk-vs-slab}, since deviation
from bulk behavior becomes more pronounced.
As expected, the sound period obtained from the bulk MD system  scales linearly
with the wavelength and matches the prediction (dash-dotted line) based on the
adiabatic sound speed from Tab.~\ref{tab:bulk_fitting_const}.

For small wavelengths, the sound period in the confined fluid follows
approximately the same linear relation as the bulk, with a slight positive
shift to longer periods---or lower sound speeds.
With increasing wavelength, deviation from bulk behavior becomes more
pronounced.
This effect becomes clearer when we plot the phase velocity directly over the
wavelength in Fig.~\ref{fig:fit_freq_bulk-vs-slab}b and compare to the
dispersion relation Eq.~\eqref{eq:speed_dispersion} derived in Sec.~\ref
{sec:theory_hans}.
As expected, the phase velocity is constant for bulk fluids in the considered
range of wavelengths.
Simulations of the confined system reveal constant speed of sound only for small
wavelengths up to approximately $120\sigma$.
For larger wavelengths, the sound speed decreases, and we found good agreement
with the dispersion relation Eq.~\eqref{eq:speed_dispersion}.
Using bulk reference data from Tab.~\ref{tab:bulk_fitting_const} and Eq.~\eqref
{eq:lambda-crit}, we predict a critical wavelength $\lambda_\mathrm
{crit}=675\sigma$.

\subsection{Critical damping and overdamped regime}

As we have described in Sec.~\ref{sec:theory_hans}, the functional form of
density and momentum correlations in confined systems change fundamentally at
the critical wavelength, when $s_\mathrm{T}$ becomes imaginary and the
trigonometric functions turn into their hyperbolic counterparts.
In the overdamped regime, the decay of sound modes is governed by two distinct
relaxation times for density and momentum correlations.
Therefore, we seek approximations to Eq.~\eqref{eq:acf_generic_long} and \eqref
{eq:acf_generic_density} which are more suitable for fitting to the available
MD data at large wavelengths.
We show in Appendix~\ref{sec:appendix_overdamped} how to arrive at effective
time correlation functions in the overdamped regime, and distinguish between
$\tau_\parallel^j$ and $\tau_\parallel^\rho$ for the relaxation times of
longitudinal momentum and density fluctuations, respectively.
Thermal relaxation is slow at large wavelengths, such that we can assume $\exp
(-D_\mathrm{T}k^2t)\approx1$ in the considered time interval.
Hence, we fitted the time autocorrelation functions of density and longitudinal
momentum in the overdamped regime to the following expressions
\begin{subequations}
    \begin{align} C_\parallel (k, t) &= \exp(-t/\tau_\parallel^j), \\ C_\rho
     (k, t) &= \frac{\gamma - 1}{\gamma} + \frac{1}{\gamma} \exp
     (-t/\tau_\parallel^\rho).
    \end{align}
\end{subequations}

\begin{figure}[!ht]
    \includegraphics[width=\columnwidth]{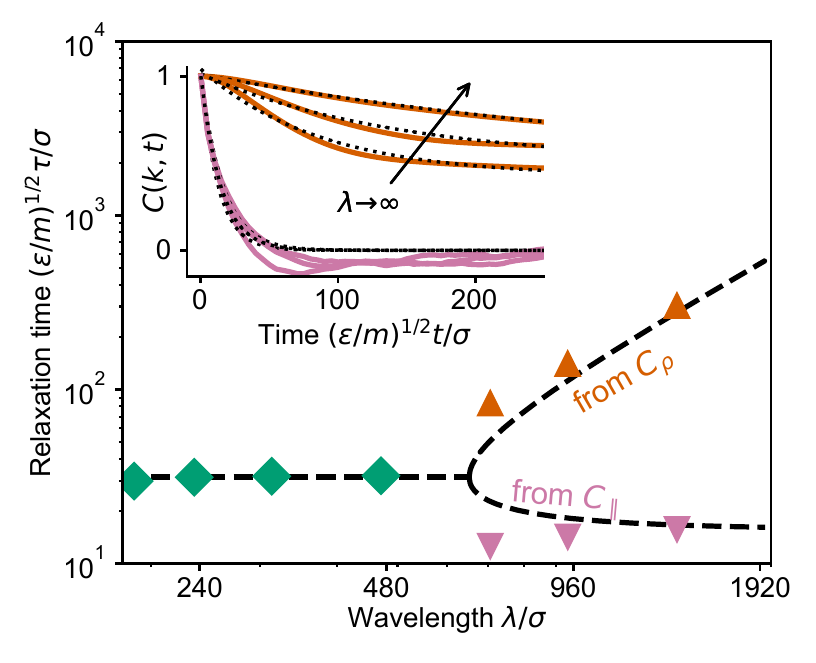}
    \caption{Transition from underdamped to overdamped dynamics. 
    The last four sound relaxation times in the underdamped regime are shown as
    green diamonds. 
    After reaching the critical wavelength, we obtain two real eigenvalues from
    Eq.~\eqref{eq:longitudinal_ev}. 
    One leads to a converging relaxation time for momentum perturbations
    (lower branch of the dashed line), the other one to a diverging relaxation
    time for density perturbations (upper branch of the dashed line). 
    For $\alpha=1.5$ and box lengths of $941.2\sigma$ and $1411.8\sigma$, we
    characterize three modes in the overdamped regime with our MD
    calculations. 
    The fits to the simplified (exponential) autocorrelation functions are shown
    in the inset, and the resulting relaxation times of density and momentum
    modes are shown as upward and downward pointing triangles, respectively.}
    \label{fig:overdamped}
\end{figure}

The critical wavelength for our confined MD simulation with gap height
$h=14.7\sigma$ and $\alpha=1.5$ is approximately $675\sigma$---more than one
half of the full box length ($941.2\,\sigma$)---such that only the largest
wavelength remains available for analysis of the overdamped relaxation.
Therefore, we further increased the box length to $L_x=1411.8\sigma$, resulting
in two more data points in the overdamped regime.
Figure~\ref{fig:overdamped} illustrates the split into two distinct relaxation
times for density and longitudinal momentum, which were formerly identical in
the underdamped region.
The dashed lines illustrate the theoretical prediction based on the eigenvalues
corresponding to longitudinal and density modes Eq.~\eqref
{eq:longitudinal_ev}.
The data points available from our MD simulations converge towards the
theoretical predictions, and the corresponding autocorrelation functions are
shown in the inset.

\subsection{Continuum simulations}
\label{sec:results_hans}

To further illustrate the suppression of sound waves in the overdamped regime,
we performed continuum simulations of confined fluids with a finite volume
implementation of Eq.~\eqref{eq:avg_balance_2d}.
The details of the implementation can be found in Ref.~\citep
{holey2022_heightaveraged}.
We studied the dynamics of density perturbations through explicit time
integration in one-dimensional slit channels with length $L_x = 1\,\mathrm
{\mu m}$ and varying gap heights between $5$ and $50\,\mathrm{nm}$.
We employed no-slip conditions at both channel walls and chose material
properties for supercritical argon, similar to our MD simulations.
The explicit time step was $\Delta t=500\,\mathrm{fs}$ and $1024$ grid cells
discretized the domain.
We initialized the system with an equilibrium density $\rho_0$ and added a
Gaussian-shaped density perturbation $\delta\rho(x)$ centered in the middle of
the channel with standard deviation $\sigma=L_x/40$.
The simulation time was short enough to avoid effects from the periodic boundary
conditions.

\begin{figure}
    \includegraphics[width=\columnwidth]{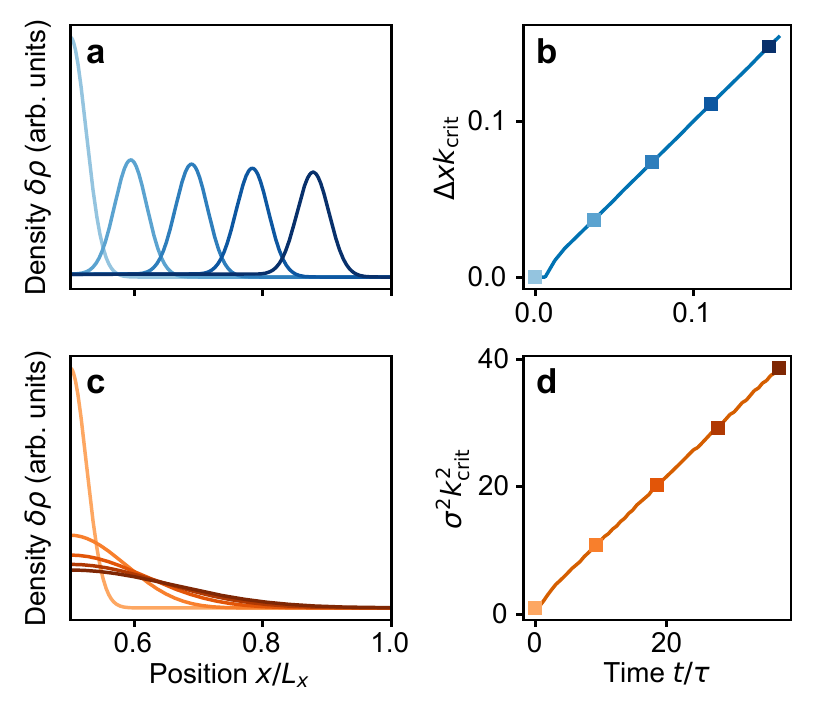}
    \caption{Time evolution of a small, Gaussian-shaped density perturbation in
     a slit geometry. The gap height in panel a is large enough to accommodate
     sound waves. Panel b shows the propagated distance $\Delta x$ over time
     $t$, rescaled by the critical wavenumber $k_\mathrm{crit}$ and the
     long-wavelength sound attenuation time $\tau =\lim_
     {k\to k_\mathrm{crit}}\tau_\parallel$, respectively. Therefore, we recover the sound
     speed $c_\mathrm{T}=1/\tau k_\mathrm{crit}$ from the slope of the curve.
     In panel c, the gap height is one order of magnitude smaller, which leads
     to a purely diffusive behavior. Panel d shows the variance $\sigma^2$ over
     time, rescaled by the squared critical wavenumber and sound attenuation
     time, respectively. Hence, the slope is proportional to the corresponding
     diffusion coefficient $D_\rho=h^2c_\mathrm{T}^2/12\nu$. Only the
     r.\,h.\,s. of the symmetric profiles is shown in a and b.}
    \label{fig:hans_density}
\end{figure}

Figure~\ref{fig:hans_density} illustrates the two limiting behaviors of sound
propagation in the underdamped regime and diffusion in the overdamped regime.
The gap height in Fig.~\ref{fig:hans_density}a--b is ten times larger than in
Fig.~\ref{fig:hans_density}c--d, while all other parameters are kept constant.
Note, that in both cases there is no single well-defined wavelength, but we take
the width of the wave package as the dominant contribution and use it to
distinguish between underdamped and overdamped behavior.

In the underdamped case, the initial Gaussian density profile separates into two
wavelets propagating to the left and right. 
Here, we show only the right half of the symmetric profiles.
Figure~\ref{fig:hans_density}b shows the propagated distance over time.
Hence, the slope tells us the speed of sound, which unsurprisingly yields
$c_\mathrm{T}$, the isothermal sound speed as defined by the initial choice of
an equation of state.

In the overdamped case, the same initial density perturbation decay is purely
diffusive.
Here, we measured the variance $\sigma^2$ of the density distribution as shown
in Fig.~\ref{fig:hans_density}d.
The linear relationship clearly illustrates the diffusive transport, and the
diffusion coefficient is given by $D_\rho=\sigma\dot{\sigma}(t)$.
The slope is unity in the normalized representation, which leads to

\begin{equation} D_\rho=\frac{1}{\tau k_\mathrm{crit}^2}=\frac{h^2c_\mathrm
 {T}^2}{12\nu}.
\end{equation}

\subsection{Effective gap height and slip length}
\label{sec:sliplength}

\begin{figure}[ht]
    \centering
    \includegraphics[width=\columnwidth]{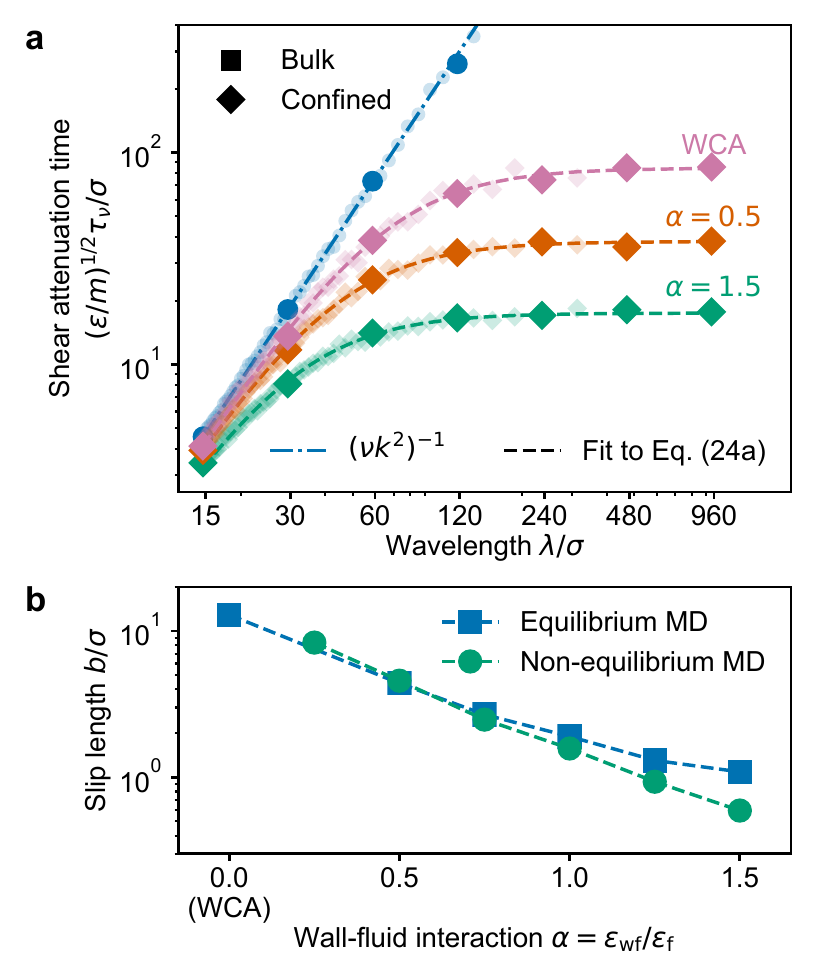}
    \caption{Shear attenuation time for varying wall-fluid interaction and
     comparison to the bulk (a). 
    The dash-dotted blue line illustrates a fit to the bulk relaxation time, and
    dashed lines show the wavelength-dependent shear relaxation time Eq.~\eqref
    {eq:relax_trans} fitted to the data from confined systems. 
    The long wavelength limit of the relaxation time determines the effective
    gap height, and since the geometric gap height is known, we can directly
    compute the slip length $b$ using the definition of $\kappa$. 
    Results of this fit procedure are shown in (b) for varying fluid wall
    interactions $\alpha$, and are compared to non-equilibrium calculations to
    obtain the slip length. 
    The dashed lines in (b) are a guide to the eyes.}
    \label{fig:shear_slip_length}
\end{figure}

Transverse momentum fluctuations do not propagate, and are therefore unaffected
by the transition at $\lambda_\mathrm{crit}$.
Due to their simple functional form and the excellent agreement of predicted and
measured relaxation times in Fig.~\ref{fig:fit_relax_bulk-vs-slab}a, we propose
the following method to quantify viscosity and slip in confined systems using
long wavelength correlations.
Instead of predicting the transition from quadratic scaling to constant
relaxation time using bulk and interfacial properties as in Sec.~\ref
{sec:results_predict}, we fit Eq.~\eqref{eq:relax_trans} to data obtained from
MD.
Thus, the characteristic transition enables us to obtain both viscosity and slip
length as fitting parameters. 
The results for such a fitting procedure for the slip length is shown in
Fig.~\ref{fig:shear_slip_length}a for various fluid-wall interaction energies.

The highest shear relaxation time can be observed when a purely repulsive
Weeks-Chandler-Anderson (WCA) potential \citep{weeks1971_role} is used for the
wall-fluid interaction.
The relaxation time decreases with increasing strength of attractive forces
between fluid and wall atoms.

With increasing wall fluid interaction, from purely repulsive to $\alpha=1.5$,
the effective viscosity increased by about $8.1\,\%$.
Compared to the bulk, the kinematic shear viscosity increased by $13.0\,\%$.
From the effective gap height, we obtain the slip length $b$ from Eq.~\eqref
{eq:slip-kappa}, and the results are shown in Fig.~\ref
{fig:shear_slip_length}b.
For comparison, we performed non-equilibrium MD simulations of identical
(but shorter) systems using moving walls.
We extracted the velocity profiles from these shear simulations by computing
time averages over slices along the gap coordinate, and calculated the slip
length from fits to the linear Couette profiles.
We obtained similar slip lengths for both methods, however, with increasing
wall-fluid interaction, the slip lengths from non-equilibrium runs tend to zero
slip, whereas the equilibrium slip lengths start to saturate at
$\alpha\geq1.0$.

\section{Discussion}
\label{sec:discussion}

Onsager's principle \citep{onsager1931_reciprocal}, the fact that small
fluctuations around an equilibrium state diffuse and propagate in the same way
as large non-equilibrium perturbations, is the foundation of several methods to
obtain transport coefficients from time correlations of collective variables at
equilibrium, with Green-Kubo methods \citep
{green1954_markoff, kubo1957_statisticalmechanical} being the most prominent
ones.
In this work, we exploit this theoretical framework to study hydrodynamic
correlations in confined systems.
In contrast to the seminal work by \citet
{bocquet1993_hydrodynamic, bocquet1994_hydrodynamic}, we considered averages
over the confining dimension and explicitly probed long wavelengths using
elongated simulation boxes.

The autocorrelation functions of density and momentum fluctuations in confined
fluids  have the same functional form as in the bulk, but transport
coefficients strongly deviate from the bulk expressions.
In a previous work, \citet{porcheron2002_propagating} came to the same
conclusion, namely that the functional form of the correlation function does
not depend on confinement, but the transport coefficients therein do.
However, they considered only pure slip boundary conditions, such that momentum
fluxes at the walls disappear.
In this case, averaging the hydrodynamic equations over the gap height leads to
a \enquote{true} two-dimensional fluid with vanishing source term in Eq.~\eqref
{eq:avg_balance_2d}.
Hence, the transition to finite relaxation times and the overdamped sound regime
do not emerge, and are difficult to observe in MD simulations if small boxes
are used, as shown in Figs.~\ref{fig:fit_relax_bulk-vs-slab} and \ref
{fig:fit_freq_bulk-vs-slab}.
Then, the interface only implicitly affects the relaxation constants, e.\,g. by
viscosity changes due to commensurability effects between the gap height and
the periodicity of fluid layering \citep{porcheron2002_propagating}.
A clear distinction between interfacial and fluid friction as recently suggested
by \citet{zhou2021_wall} is therefore difficult, when ordering effects reach
far into the fluid region and confining dimensions are small.

Ordering effects are small in the presented MD simulations of confined LJ
fluids, and therefore prediction of the spectral relaxation behavior based on
bulk fluid properties works well in most situations.
The prediction works particularly well for transverse modes, as shown in
Fig.~\ref{fig:fit_relax_bulk-vs-slab}a, where the shear viscosity and slip
length has been obtained from reference simulations in equivalent MD setups.
In fact, fits to transverse momentum autocorrelation functions as proposed
by \citet{palmer1994_transversecurrent} is an established method to calculate
the viscosity at equilibrium \citep{hess2002_determining} and has been recently
extended by \citet{cheng2020_computing} to compute heat conductivity from
density correlations.

Quantification of slip in MD simulations is of great importance for the
understanding of confined fluid systems, e.\,g. for multiscale simulations of
fluid transport \citep{holland2015_molecular} or lubrication \citep
{savio2015_multiscale}.
The characteristic transition from bulk-like to wavelength-independent
relaxation offers a new path to determine the slip length at equilibrium, as
illustrated in Fig.~\ref{fig:shear_slip_length}.
The proposed method is in the spirit of the early work by \citet
{palmer1994_transversecurrent} and has the advantage that both interfacial slip
and the shear viscosity of the fluid can be determined simultaneously.
This can be seen as extension to a similar approach by \citet
{sokhan2008_slip}, where viscosity is an input parameter, that has to be taken
from literature, or is obtained from separate MD simulations.

Figure~\ref{fig:shear_slip_length}a highlights that even for purely repulsive
fluid-wall interactions relaxation times eventually converge to a finite limit
with increasing wavelength.
Hence, the limit of infinite slip, where Galilean invariance is fully restored,
is impossible to reach with MD simulations of interacting fluid and wall
atoms.
However, in the case of ultra-low interfacial friction, it requires large
wavelengths to observe deviations from a bulk system which are usually not
probed in conventional MD systems.

The prediction of the slip length based on the equilibrium simulations have been
compared with results from non-equilibrium calculations in Fig.~\ref
{fig:shear_slip_length}b.
Slip lengths from both calculations decay exponentially with increasing wall
fluid interaction.
Equilibrium results seem to converge towards a finite slip length of
approximately one atomic diameter, which is probably the closest one can get to
a no-slip boundary condition in an atomistic system.
This agrees with other equilibrium slip length measurements \citep
{huang2014_greenkubo}, suggesting that friction converges and leads to a small
but finite slip length, even for strong fluid-wall interaction.
Non-equilibrium measurements rely on linear extrapolation of the velocity
profiles to obtain the slip length, which is very sensitive to the exact
location of the fluid-wall interface.
This might explain the deviation of the non-equilibrium results at higher
interaction energies.
For practical use of the proposed method, shorter box sizes on the order of a
few gap heights might be sufficient, which drastically reduces the
computational effort compared to the MD setup shown here.

Longitudinal momentum relaxation times show a similar behavior as in the
transverse direction, as shown in Fig.~\ref{fig:fit_relax_bulk-vs-slab}b.
Yet, in contrast to the transverse case, sound absorption depends on viscous and
thermal effects.
Therefore, predicted relaxation times for the bulk system based on literature
data and our own simulation are  slightly above the ones obtained from the
autocorrelation functions.
The isothermal theory for confined systems accurately describes the long
wavelength relaxation, since the time required for heat to diffuse along the
lateral direction is long enough to assume isothermal conditions.
However, at short wavelengths, the theory overestimates relaxation times, but
the prediction can be improved by considering thermal effects in the bulk
contribution to the overall relaxation.
Remaining deviations between our prediction and the data, might be due to
viscosity enhancement in the confined system compared to the bulk due to
ordering effects introduced by the walls, as discussed above.

The role of thermal effects in the presented MD results is highlighted in
Fig.~\ref{fig:fit_relax_bulk-vs-slab}c.
Relaxation times for the Rayleigh process contribution in the density
autocorrelation function obtained from bulk and confined systems are
indistinguishable, since rigid walls are used.
In this context, rigid walls imply that there is no thermal coupling between
fluid and wall atoms, since wall atoms do not vibrate, i.\,e. there is no heat
flux across the interface.
In analogy to wall slip described above, this represents an idealized situation
with a perfect \enquote{thermal slip} condition, i.\,e. an interface with
infinite Kapitza length \citep{barrat2003_kapitza}.
Therefore, similar transition to finite relaxation times is expected for systems
with vibrating walls, which is shown in Fig.~\ref
{fig:fit_relax_bulk-vs-slab}c, but is not the main focus of this paper.

Instead, we focus on the long-wavelength behavior of propagating modes.
The dispersion relation, Eq.~\eqref{eq:speed_dispersion}, is dominated by the
longitudinal relaxation time in the long wavelength limit and therefore
independent of thermal effects.
Hence, considering vibrating walls does not affect the existence of the
transition to overdamped sound.
As shown in Fig.~\ref{fig:fit_freq_bulk-vs-slab}, both sound period and
accordingly the velocity of sound in confined fluids can be adequately
described with Eq.~\eqref{eq:speed_dispersion}.
In bulk fluids, dispersion is usually observed at molecular lengths scales,
where transport coefficients depend on wavelength, and short-term memory
effects have to be considered within a generalized hydrodynamic theory \citep
{boon1980_molecular}.
In confined systems, dispersion is an intrinsic property of the system, where
both fluid and interfacial properties determine the lengthscale at which it
becomes effective.

The predicted bifurcation into diverging density and converging momentum
relaxation times in the overdamped regime has been confirmed by our simulations
in very large systems, as shown in Fig.~\ref{fig:overdamped}.
Hence, long wavelength modes in confined systems are transported entirely by
diffusion.
The short-lived longitudinal momentum correlations can be interpreted as
discrete \enquote{jumps} of a wave package, that bring about diffusive mass
transport in hydrodynamic systems with additional dissipation due to walls,
similar to Fickian diffusion for discrete particles.
The size of typical MD simulations of confined fluids does not probe this limit,
although it might be the dominant transport mechanism in real confined
systems.
For instance, \citet{cheng2002_fluid} measured slip lengths of $25\,\mathrm
{nm}$ in a $50\,\mathrm{nm}$ thin channel for hexadecane on photoresist-coated
glass.
The critical wavelength for this system is on the order of micrometers, thus
reaching into the frequency range of ultrasound applications.

The results of our continuum simulations shown in Fig.~\ref
{fig:hans_density} illustrate the effect of overdamped sound in a
non-equilibrium scenario.
By reducing the gap height, we interpolate between sound propagation and
diffusion, as initially suggested by \citet{ramaswamy1982_linear}.
The empirically determined diffusion coefficient agrees with the one reported in
Ref.~\citep{pagonabarraga1999_shorttime}.
Transition to overdamped behavior may also lead to arrest of an initially
propagating wave package due to spread related to dissipation in the
underdamped regime or due to spatially varying gap height or wall slip.
The latter suggests further work on the role of roughness on the critical
transition of sound modes in confined systems.

\section{Conclusion}

In this work, we showed that correlations of the hydrodynamic conserved
variables in confined fluids can be derived from an isothermal height-averaged
description of continuum balance equations.
The functional form of hydrodynamic correlation functions remains equivalent to
the bulk, but characteristic time scales therein are affected by confinement.
We focused on the lateral wavelength dependence of relaxation times and phase
velocities.
The continuum description predicts a transition to constant relaxation times of
density and momentum perturbations, which we confirmed by upscaling MD
simulations to the long wavelength limit.
Furthermore, our theory contains a geometry-dependent dispersion relation for
the speed of sound in the long wavelength limit, which leads to a transition
from underdamped to overdamped dynamics, which is also evident from the MD
simulations.
Large MD boxes are required to probe the overdamped limit, but the transition
can be on the order of the system size in highly confined fluids.
Hence, diffusive contributions to lateral mass transport might be systematically
neglected in finite systems.
Finally, we proposed a new equilibrium method to calculate the slip length in
confined MD systems based on our findings, which shows accurate results when
compared to a non-equilibrium reference.

\begin{acknowledgments} The authors gratefully acknowledge support by the German
 Research Foundation (DFG) through GRK 2450. Furthermore, the authors
 acknowledge support by the state of Baden-Württemberg through bwHPC, for
 calculations carried out on bwForCluster NEMO (DFG grant INST 39/963-1 FUGG)
 and bwUniCluster2.0. Data is stored on bwSFS (University of Freiburg, Deutsche
 Forschungsgemeinschaft Grant No. INST 39/ 1099-1 FUGG).
\end{acknowledgments}

\appendix

\section{Effective gap height with slip}\label{sec:appendix-kappa} The effective
 gap height $h^\ast=\sqrt{\kappa}h$ can be obtained from Eq.~\eqref
 {eq:equal_flux}, with a parabolic slip velocity profile
\begin{equation}\label{eq:vprofile_slip} u(z) = az(z - h) + \frac{U_2 - U_1}
 {h}z + U_1,
\end{equation}
where $U_1$ and $U_2$ are the slip velocities at the bottom and top wall
respectively. The corresponding no-slip velocity profile is
\begin{equation} u^\ast(z) = az(z - h^\ast).
\end{equation}

The definition of the Navier slip length relates $U_1$ and $U_2$ with $b_1$ and
$b_2$ through
\begin{equation}
    \begin{split} U_1 &= u'(0) b_1, \\ U_2 &= -u'(h) b_2,
    \end{split}
\end{equation} such that we can substitute
\begin{equation}
    \frac{U_2 - U_1}{h} = a h \frac{b_1 - b_2}{h + b_1 + b_2},
\end{equation} and
\begin{equation} U_1 = \left(\frac{b_1 - b_2}{h + b_1 + b_2} - 1\right)a h b_1
\end{equation} in Eq.~\eqref{eq:vprofile_slip}.
Evaluating Eq.~\eqref{eq:equal_flux} eventually leads to
\begin{equation} h^{\ast2} = \underbrace{\left[1 +\frac{6b_1}{h} - 3\frac
 {b_1-b_2}{h+b_1+b_2}\left(1 + \frac{2b_1}{h}\right)\right]}_{=:\kappa} h^2.
\end{equation}

\section{Imaginary parts}
\label{sec:appendix_imag}

The imaginary parts of the solution to Eq.~\eqref{eq:isoT_2D_ode} are shown here
for completeness.

\begin{equation}
\begin{split}
\mathrm{Im}[\rho(k,t)] &= -\frac{j_\parallel(k, 0)}{s_\mathrm{T}} e^{-\frac{t}
 {\tau_\parallel}}
\sin(s_\mathrm{T} k t), \\
\mathrm{Im}[j_\parallel(k,t)] &=
- \rho(k, 0) e^{-\frac{t}{\tau_\parallel}}
\left(s_\mathrm{T} - \frac{1}{s_\mathrm{T} \tau_\parallel^2 k^2} \right) \sin
 (s_\mathrm{T} k t).
\end{split}
\end{equation}

\section{Effective relaxation for overdamped dynamics}
\label{sec:appendix_overdamped}

The limiting behavior of Eq,~\ref{eq:acf_slab_underdamped} in the overdamped
regime is easily obtained by taking the limit
\begin{equation}
    \lim_{k\to0} \iu s_\mathrm{T} k = \lim_{k\to0} - \sqrt{\left(\frac{6\nu}
     {h^2\kappa}\right)^2 - c_\mathrm{T} k^2} = -\frac{6\nu}{h^2\kappa}.
\end{equation}
Hence, the normalized Fourier coefficients of longitudinal momentum fluctuations
are identical to the transverse ones
\begin{equation}
    \begin{split}
    \lim_{k\to0} \tilde{j}_\parallel(k,t) = &\exp\left(-\frac{6\nu}
     {h^2\kappa}\right)\times \\
    &\left[\cosh\left(-\frac{6\nu}{h^2\kappa}\right) + \sinh\left(-\frac{6\nu}
      {h^2\kappa}\right)\right] \\ = &\exp\left(-\frac{12\nu}
      {h^2\kappa}\right) = \tilde{j}_\perp(k,t).
    \end{split}
\end{equation}
However, the normalized Fourier coefficients of density fluctuations become
\begin{equation}
    \begin{split}
    \lim_{k\to0} \tilde{\rho}(k,t) = &\exp\left(-\frac{6\nu}
     {h^2\kappa}\right)\times \\
    &\left[\cosh\left(-\frac{6\nu}{h^2\kappa}\right) - \sinh\left(-\frac{6\nu}
      {h^2\kappa}\right)\right] = 1,
    \end{split}
\end{equation}
and we are now interested how this limit is reached.
For the sake of brevity, we write $a\equiv6\nu/h^2\kappa$ and $b \equiv -\iu
s_\mathrm{T} = \sqrt{(a/k)^2-c_\mathrm{T}^2}$ with $c_\mathrm{T}<a/k$, i.\,e.
\begin{equation}
    \tilde{\rho}(k,t) = e^{-at}\left[\cosh(b k t) + \frac{a}{b k}\sinh(b k
     t)\right],
\end{equation} where we have used the symmetries of the hyperbolic functions. At
 times $t_0$ much larger than a characteristic time of the system
 ($t_0\gg1/bk$), we can assume $\sinh(b k t_0) = \cosh(b k t_0)$. Expanding
 $\tilde{\rho(k, t)}$ around $t_0$ leads to

\begin{equation}\label{eq:dens_od_expansion}
    \begin{split}
    \tilde{\rho}(k, t) &= \tilde{\rho}(k, t_0) + \frac{b k + a}{b k}e^
     {-a t_0}\cosh(b k t_0)\times \\
    &\left[(b k - a)(t-t_0) + \frac{1}{2!}(b k - a)^2
      (t-t_0)^2 + \cdots\right]\\
    &= \tilde{\rho}(k, t_0)\times \\
    &\left[1 + (b k - a)(t-t_0) + \frac{1}{2!}(b k - a)^2
      (t-t_0)^2 + \cdots\right]\\
    &= \tilde{\rho}(k, t_0)e^{(b k - a)(t-t_0)}.
    \end{split}
\end{equation}

\begin{figure}[!ht]
    \includegraphics[width=\columnwidth]{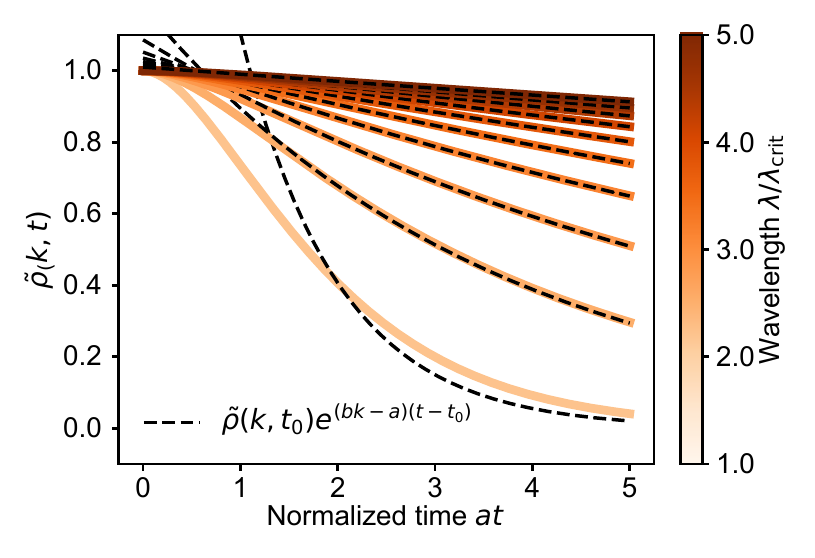}
    \caption{Same as Fig.~\ref{fig:slab_theory}c, but with the effective
     long-time expression for $\tilde{\rho}(k,t)$. The expansion shown in
     Eq.~\ref{eq:dens_od_expansion} is around $t_0=2/a$, and except for the
     critically damped mode, we obtain a good agreement with the original form
     for large $t$.}
    \label{fig:appendix_dens_overdamped}
\end{figure}

At long times, density modes in the overdamped regime decay exponentially in
time with a rate $b k - a < 0$, where $a$ is a geometry-dependent constant and
$b k - a \sim \mathcal{O}(k^2)$ as shown in Fig.~\ref
{fig:appendix_dens_overdamped}.

\FloatBarrier

\end{document}